\documentclass[12pt, leqno]{article}
\usepackage{amsmath,setspace,multirow,lineno}
\usepackage[font=small,labelfont=bf]{caption}
\usepackage[top=1.1in, bottom=1.1in, left=1.1in, right=1.1in]{geometry}

\usepackage[round]{natbib}
\usepackage{color,soul}

\DeclareCaptionStyle{italic}[justification=centering]{labelfont={bf},textfont={it},labelsep=colon}
\usepackage{graphicx,psfrag,epsf,textcomp,epstopdf, amsthm}
\usepackage{enumerate, dsfont}
\usepackage{natbib}
\usepackage{url,xcolor}
\usepackage{booktabs, subfig, bm, paralist,mathpazo,tikz,todonotes,longtable,microtype}
\usepackage[linesnumbered,ruled,vlined]{algorithm2e}

\usepackage[pdftex,colorlinks=true]{hyperref}
\definecolor{darkblue}{rgb}{0,0,.6}
\hypersetup{citecolor=darkblue,linkcolor=darkblue,urlcolor=darkblue}
\definecolor{DarkRed}{rgb}{.7,0,.4}

\newcommand{\argmax}{\operatornamewithlimits{argmax}}

\usepackage{comment,orcidlink}

\newcommand{\blind}{0}

\newcommand{\X}{\mathcal{X}}

\newcommand{\Rlogo}{\protect\includegraphics[height=1.8ex,keepaspectratio]{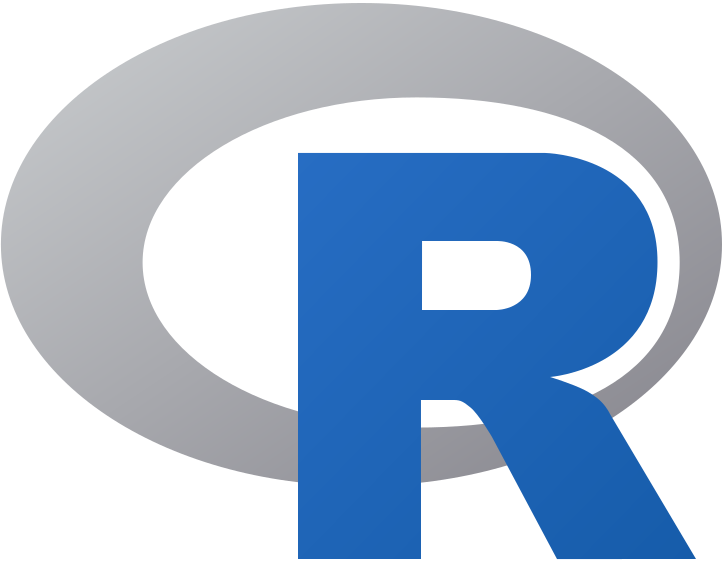}}

\graphicspath{{plots/}}
\DeclareMathOperator*{\argmin}{\arg\!\min}

\newsavebox\CBox

 \newtheorem{@definition}{\sc Definition}[section]

  \renewcommand\X{\mathcal{X}}

\date{}

\begin{document}

\def\spacingset#1{\renewcommand{\baselinestretch}{#1}\small\normalsize} \spacingset{1}

\if0\blind
{
\title{\bf A Robust Functional Partial Least Squares for Scalar-on-Multiple-Function Regression}}
\author{
Ufuk Beyaztas\footnote{Corresponding address: Department of Statistics, Marmara \"{U}niversitesi G\"{o}ztepe Yerle\c{s}kesi, 34722 Kadik\"{o}y-Istanbul, Turkey; Email: ufuk.beyaztas@marmara.edu.tr}  \orcidlink{0000-0002-5208-4950}
\\
Department of Statistics \\
Marmara University \\
\\
Han Lin Shang \orcidlink{0000-0003-1769-6430} \\
    Department of Actuarial Studies and Business Analytics \\
    Macquarie University 
}
\maketitle
\fi

\if1\blind
{
\title{\bf A Robust Functional Partial Least Squares for Scalar-on-Multiple-Function Regression}
} \fi

\maketitle

\begin{abstract}
The scalar-on-function regression model has become a popular analysis tool to explore the relationship between a scalar response and multiple functional predictors. Most of the existing approaches to estimate this model are based on the least-squares estimator, which can be seriously affected by outliers in empirical datasets. When outliers are present in the data, it is known that the least-squares-based estimates may not be reliable. This paper proposes a robust functional partial least squares method, allowing a robust estimate of the regression coefficients in a scalar-on-multiple-function regression model. In our method, the functional partial least squares components are computed via the partial robust M-regression. The predictive performance of the proposed method is evaluated using several Monte Carlo experiments and two chemometric datasets: glucose concentration spectrometric data and sugar process data. The results produced by the proposed method are compared favorably with some of the classical functional or multivariate partial least squares and functional principal component analysis methods.
\end{abstract}


\newpage
\spacingset{1.5} 

\section{Introduction} \label{sec:intro}

Functional data are collected over a continuum such as time, spatial grids, and wavelengths. The need for developing reliable statistical techniques to analyze this kind of data has been increased in many scientific branches, including chemometrics. The interested readers are referred to \cite{ramsay1991}, \cite{ramsay2002, ramsay2006}, \cite{ferraty2006}, \cite{horvath2012}, \cite{cuevas2014}, \cite{Hsing}, and \cite{KoRe} for theoretical results and applications of functional data analysis methods.

Spectroscopic techniques are the most commonly used instrumental techniques in chemometrics analysis of the data obtained in many applied sciences such as herbal medicine \citep{Ankit2014} and biological systems \citep{Skolik2018}. They are fast, precise, and reproducible \citep{Kamila2020}. Chemometrics analysis of the data obtained from spectroscopy devices resulting from analysis with spectroscopic techniques (called spectral data) helps extract information about the sample's chemical composition and make statistical inferences about it. Spectral data are the absorbance spectrum of chemicals and can be seen as a function of the wavelength; thus, it is natural to work with functional data in chemometrics. Consult \cite{Saeys2008} and \cite{Aguilera2013} for many applications of functional data analysis in chemometrics. Since \cite{Cardot1999}, the scalar-on-function regression model, whose response variable is scalar and predictors consist of random curves, has been well studied in chemometrics analysis of spectral data \citep[see e.g.,][]{Ferraty2002, Reiss2007, Matsui2008, Jiang2008, Kramer2008, Aguilera2010, Aguilera2016, Lopez2017, Smaga2018}. The numerical results of these papers have been shown that the functional regression models perform better than traditional multivariate statistical techniques in modeling spectral data, thanks to the continuity and smoothness of the functional data.

Let us consider a random sample $\left\lbrace Y_i, \X_i(t): i = 1, \ldots, n \right\rbrace$ from the pair $( Y, \X )$, where $Y \in \mathbb{R}$ is a scalar random variable and $\X = \left\lbrace \X(t) \right\rbrace_{t \in [0, \mathcal{I}]}$ is a stochastic process whose elements are represented by curves belonging to $\mathcal{L}_2$ Hilbert space, denoted by $\mathcal{H}$, with bounded and closed interval $t \in \mathcal{I}$. We assume that $\X(t)$ is a $\mathcal{L}_2$ continuous stochastic process. In addition, without loss of generality, we assume that both $Y$ and $\X(t)$ are mean-zero processes, so that $\text{E}[Y] = \text{E}[\X(t)] = 0$, and $t \in [0,1]$. Then, the scalar-on-function regression model has the following form:
\begin{equation} \label{eq:sof}
Y_i = \int_0^1 \X_i(t) \beta(t) dt + \epsilon_i, 
\end{equation}
where $\beta(t) \in \mathcal{L}_2[0,1]$ is the regression coefficient function, and $\epsilon_i$ is a stochastic error term, which is generally assumed to be independent and identically distributed Gaussian real random variable with mean-zero and variance $\sigma^2$. In addition, the error term $\epsilon_i$ is assumed to be independent of $\X_i(t)$.

In Model~\eqref{eq:sof}, the relationship between the scalar $Y$ (called response) and $\X(t)$ (called functional predictor) is characterized with the functional regression coefficient $\beta(t)$. However, the estimation of Model~\eqref{eq:sof} is a difficult task as $\beta(t)$ belongs to an infinite-dimensional space by its nature. The common practical approach to overcome this vexing issue is based on the dimension reduction paradigm. With this approach, the functional regression coefficient $\beta(t)$ is projected onto a finite-dimensional space via a dimension reduction technique. 

Several basis expansion methods have been developed:
\begin{inparaenum}
\item[1)] The fundamental basis functions include $B$-spline, Fourier, and wavelet basis, \citep[see, e.g.,][]{Marx1999, Cardot2003, CaiHall, Goldsmith2011, Zhao2012}. However, such basis function expansion methods may require a large number of basis functions to approximate the functional regression coefficient, causing overfitting of the model and low prediction accuracy.
\item[2)] Functional principal component (FPC) regression uses the finite-dimensional projections of the functional objects to estimate Model~\eqref{eq:sof}. The principal components are constructed using the covariance function of the functional predictors. Compared with the general basis functions, FPC is more informative since it is data-driven and uses the information of functional predictors gathered from their covariance functions \citep[see, e.g.,][]{Yao2007, Hall2007, Lee2012, Reiss2017, Wang2016, Goldsmith2014}. On the other hand, a few FPCs usually contain most of the information related to the functional predictors' covariance, and these are not necessarily important for representing FPCs. All or some of the most important terms accounting for the interaction between the basis functions and functional predictors might come from later principal components \citep{Delaigle2012}.
\item[3)] When extracting the latent components, functional partial least squares (FPLS) regression uses both response and predictor variables. Compared with the FPC, the FPLS can capture the relevant information with fewer terms, and thus, it is generally preferred to the FPC \citep[see, e.g.,][]{Reiss2007, Delaigle2012, Bande2017, Yu2016}. 
\end{inparaenum}

In chemometrics data analyses, the use of the FPLS leads to better prediction of the response variable than the FPC since FPLS considers the information of the response variable when extracting the FPLS components. Furthermore, \cite{Aguilera2010} showed that the FPLS provides better estimations of parameter function for the Model~\eqref{eq:sof} than the one obtained from the FPC. Therefore, we consider FPLS regression in estimating Model~\eqref{eq:sof}.

FPLS components are obtained by maximizing the squared covariance between the scalar response $Y$ and functional predictor $\X(t)$. The least-squares (LS) loss function, whose solution yields the LS estimator, optimizes the covariance operator used for this purpose. However, the LS estimator is susceptible to outliers, far from the bulk of the data. In such a case, the extracted FPLS components may be unreliable, and they may not reflect the underlying structure of the data. Consequently, the FPLS method may produce biased parameter estimations, leading to a poor prediction of the response variable. Several approaches have been proposed to robustly estimate the scalar-on-function regression model~\eqref{eq:sof} in the presence of outliers. For example, \cite{Maronna} proposed a robust version of a spline-based estimate by replacing a bounded loss function with an LS loss function. \cite{ShinLee} presented a robust procedure using outlier-resistant loss functions, where the robust estimates are computed via an iteratively reweighted penalized LS algorithm. \cite{Kalogridis} proposed a two-step estimation procedure combining robust functional principal components and robust linear regression. However, to our knowledge, there is no approach for Model~\eqref{eq:sof} to extract its FPLS components and produce a reliable prediction robustly. Further, existing methods only estimate the simple scalar-on-function regression model, where there is only one functional predictor in the model \citep[see, e.g.,][]{ShinLee, Kalogridis}.

We propose a robust FPLS method, abbreviated as RFPLS, which allows more than one functional predictor in the model, to obtain a robust estimation for Model~\eqref{eq:sof}. We construct our robust approach extending the partial robust M-regression algorithm of \cite{Serneels2005} to the functional data. Our proposed method consists of two steps. In the first step, the RFPLS uses a weighted covariance operator to extract FPLS components in the presence of outliers robustly. In the second step, the M-estimator and Tukey's bisquare loss function are used to solve the regression problem of scalar response on the extracted FPLS components and to approximate the regression coefficient function of Model~\eqref{eq:sof}. Compared with LS loss function, bisquare loss function varies more slowly at large values. On the other hand, it sacrifices some efficiency for achieving a robustness degree \citep{Jiang2019}. To overcome this problem, we use the data-driven approach of \cite{Wang2007, Wang2018} to determine the optimum tuning parameter, which allows achieving the desired robustness level without sacrificing efficiency. 

Generally, there are two types of outliers in a regression problem, leverage points (outliers in the predictor space) and vertical outliers (outliers in the response variable). In an FPLS regression, the effects of leverage points are reflected by the FPLS components. In the first step of the proposed RFPLS method, the effects of leverage points are down-weighted by the weighted covariance operator. In the second step, on the other hand, the effects of vertical outliers are down-weighted by the M-estimator. Therefore, the proposed RFPLS method is robust to leverage points in the predictor variables and vertical outliers in the response variable.

The functional predictors belong to an infinite-dimensional space. However, in practice, the random sample curves are observed in a finite set of discrete-time points. Thus, the calculation of FPLS components becomes problematic because they need to be calculated using discretely observed sample curves. The functional form of discretely observed data is first approximated via a finite-dimensional basis expansion method to overcome this problem. Then, the FPLS regression model constructed based on the infinite-dimensional random variables is approximated via the multivariate PLS regression constructed using the basis expansion coefficients \citep[see, e.g.,][]{Aguilera2010, Aguilera2016}. In this paper, the $B$-spline basis function expansion approximates the functional forms of the discretely observed data and approximates the FPLS components. The proposed method's predictive performance depends on the number of FPLS components. To this end, we apply a mean squared prediction error (MSPE)-based k-fold cross-validation procedure on determining the optimum number of FPLS components.

The remaining part of this paper is organized as follows. Sections~\ref{sec:methodology} and~\ref{sec:rfpls} summarize the FPLS and proposed RFPLS methods, respectively. In Section~\ref{sec:results}, several Monte Carlo experiments, as well as two empirical data analyses, are performed to evaluate the finite-sample performance of the proposed method. Section~\ref{sec:conc} concludes the paper, along with some ideas on how the methodology can be further extended. The technical details of the FPLS and proposed RFPLS methods are given in Appendix~\ref{sec:app_A} and~\ref{sec:app_B}, respectively.

\section{Functional PLS regression} \label{sec:methodology}

Let $Y \in \mathbb{R}$ and $\left\lbrace \X_m(t) \right\rbrace_{1 \leq m \leq M}$ respectively denote a scalar response and $M$ set of functional predictors consisting of random functions, where $\X_m(t) \in \mathcal{L}_2[0,1]$, $\forall~m = 1, \ldots, M$ with respect to a Lebesgue measure $dt$ on $[0,1]$. Denote by $\bm{\X}(t) = [ \X_{1}(t), \ldots, \X_{M}(t) ]^\top$ the vector in a Hilbert space of $M$-dimensional  vectors of functions and $^{\top}$ denotes a vector transpose. Let $\mathcal{H}^M = \mathcal{H} \oplus \ldots \oplus \mathcal{H}$ denote the direct sum of the $M$ seperable Hilbert spaces, then $\bm{\X}(t) \in \mathcal{H}^M = \mathcal{L}_2^M[0,1]$ \citep[e.g.,][]{reed1980}. We assume that $\bm{\X}(t)$ is a $\mathcal{L}_2$ continuous stochastic process, which implies that $\X_m(t)$ is a $\mathcal{L}_2$ continuous stochastic process, $\forall~ m = 1, \ldots, M$. Further, without loss of generality, we assume that both $Y$ and $\X_m(t)$, for $m = 1, \ldots, M$, are mean-zero processes, so that $\text{E}[Y] = \text{E}[\X_m(t)] = 0$. Then, the scalar-on-multiple-function regression model between $Y$ and $\bm{\X}(t)$ has the following form:
\begin{align}
Y_i &= \sum_{m=1}^M \int_0^1 \X_{im}(t) \beta_m(t) dt + \epsilon_i, \nonumber \\
&= \int_0^1 \bm{\X}_i^\top(t) \bm{\beta}(t) dt + \epsilon_i, \label{eq:msof}
\end{align}
where $\bm{\X}_i(t) = [ \X_{i1}(t), \ldots, \X_{iM}(t) ]^\top$, $\beta_m(t) \in \mathcal{L}_2[0,1]$ is the regression coefficient function linking $Y$ with $\X_m(t)$, and $\bm{\beta}(t) = [ \beta_1(t), \ldots, \beta_M(t) ]^\top \in \mathcal{L}_2^M[0,1]$.

The FPLS components of Model~\eqref{eq:msof}, denoted by $\bm{\xi}$, are obtained as linear functionals of $\bm{\X}(t)$ \citep{Aguilera2016},
\begin{equation*}
\bm{\xi} = \int_0^1 \bm{\X}^\top(t) \bm{w}(t) dt,
\end{equation*}
as solutions of the Tucker's criterion \citep{Tucker}, i.e., $\text{Cov}^2 \left( Y, \int_0^1 \bm{\X}^\top(t) \bm{w}(t) dt \right)$, where $\bm{w}(t) = [ w_1(t), \ldots, w_M(t) ]^\top \in \mathcal{L}_2^M[0,1]$ is the vector of eigenfunctions associated to the FPLS component $\bm{\xi} = [ \xi_1, \ldots, \xi_M ]^\top$. Let $\mathcal{C}_{Y \bm{\X}}$ and $\mathcal{C}_{\bm{\X} Y}$ respectively denote the cross-covariance operator, which evaluates the contribution of $M$-variate functional predictor $\bm{\X}(t)$ to the scalar response $Y$, and its adjoint. Then, the spectral analysis of $\mathcal{U} = \mathcal{C}_{\bm{\X} Y} \circ \mathcal{C}_{Y \bm{\X}}$ leads to a countable set of positive eigenvalues $\lambda$ associated to an $M$-variate orthonormal basis of eigenfunctions $\bm{w} = [ w_{1}, \ldots, w_{M} ]^\top \in \mathcal{L}_2^M[0,1]$ as a solution of
\begin{equation}\label{eq:spec}
\mathcal{U} \bm{w} = \lambda \bm{w},
\end{equation}
where $\int_0^1 \sum_{m=1}^M w_{m}(t) w_{m}^\top(t) dt = 1$.

Denote by $\bm{w}^{(1)}(t)$ the eigenfunction of $\mathcal{U}$ associated to its largest eigenvalue. Then, the first FPLS component, denoted by $\bm{\xi}^{(1)}$, is computed as follows:
\begin{equation*}
\bm{\xi}^{(1)} = \int_0^1 \bm{\X}^\top(t) \bm{w}^{(1)}(t) dt.
\end{equation*}
The subsequent FPLS components are obtained iteratively, where at each iteration, the FPLS component is obtained by taking into account the information gathered from the previous iteration as follows:
\begin{equation*}
\bm{\xi}^{(h)} = \int_0^1 [\bm{\X}^{(h-1)}(t)]^{\top} \bm{w}^{(h)}(t) dt,
\end{equation*}
where $\bm{w}^{(h)}(t)$ and $\bm{\X}^{(h-1)}(t)$ denote the weight function and the residual obtained at step $h$, respectively.

\subsection{Regression coefficient estimation via basis expansion}\label{sec:bee}

The functional random variables belong to an infinite-dimensional space, however in practice, the values of functional random variables are observed in a finite set of discrete time points $\X_m(t_{mj});$ $0=t_{m1} < ... < t_{mJ_m}=1 $. Thus, the FPLS components cannot be calculated directly from the discretely observed data. To overcome this problem, similar to \cite{Aguilera2010, Aguilera2016} and \cite{Julien2014}, we consider approximating the functional forms of the predictors via a basis expansion method from the discretely observed data. It reduces the problem of eigen-analysis in~\eqref{eq:spec} to the finite-dimensional eigen-analysis problem. Suppose that each curve $\X_{im}(t)$ for $i = 1, \ldots, n$ and $m = 1, \ldots, M$, admits the basis expansion
\begin{equation*}
\X_{im}(t) = \sum_{k=1}^{K_m} d_{imk} \psi_{mk}(t),
\end{equation*}
where $K_m$ is the number of basis functions for $m^\textsuperscript{th}$ functional predictor, $\psi_{mk}(t)$ is a basis in $\mathcal{L}_2[0,1]$, and $\bm{d}_{imk}$ denotes the corresponding basis expansion coefficient. Let $\bm{D}$ denote the $n \times \sum_{m=1}^M K_m$-dimensional matrix with row entries $\bm{d}_i = [ d_{i11}, \ldots, d_{i1K_1}, d_{i21}, \ldots, d_{i2K_2}, \ldots, d_{iM1}, \ldots, d_{iMK_M} ]$. Let also $\bm{\psi}(t)$ denote the block-diagonal matrix with entries $\bm{\psi}_m(t) = [ \psi_{m1}(t), \ldots, \psi_{m K_m} ]^\top$ for $m = 1, \ldots, M$ as follows: 
\begin{equation*}
\bm{\psi}(t) = \begin{bmatrix} 
\bm{\psi}_1^\top(t) & \bm{0} & \cdots & \bm{0} \\
\bm{0} & \bm{\psi}_2^\top(t) & \cdots & \bm{0} \\
\cdots & \cdots & \cdots & \cdots \\
\bm{0} & \bm{0} & \cdots & \bm{\psi}_M^\top(t)
\end{bmatrix}
\end{equation*}
Then, $\bm{\X}(t)$ has the following basis expansion form
\begin{equation*}
\bm{\X}(t) = \bm{D} \bm{\psi}^\top(t).
\end{equation*} 

From~\eqref{eq:spec}, the eigenfunctions and therefore the $M$-variate parameter function in Model~\eqref{eq:msof} can be expressed as a linear combination of $\bm{\psi}(t)$ and the corresponding basis expansion coefficients as follows:
\begin{equation*}
\bm{w}(t) = \bm{\psi}(t) \bm{w}, \qquad \bm{\beta}(t) = \bm{\psi}(t) \bm{\beta},
\end{equation*}
where $\bm{w} = [ w_{11}, \ldots, w_{1K_1}, w_{21}, \ldots, w_{2K_2}, \ldots, w_{M1}, \ldots, w_{MK_M} ]^\top$ and $\bm{\beta} = [ \beta_{11}, \ldots, \beta_{1K_1}, \beta_{21}, \ldots, \beta_{2K_2}, \ldots,\\ \beta_{M1}, \ldots, \beta_{MK_M} ]^\top$ denote the vectors of basis expansion coefficients. Let $\bm{\Psi} = \int_0^1 \bm{\psi}(t) \bm{\psi}^\top(t) dt$ denote the symmetric block-diagonal $\sum_{m=1}^M K_m \times \sum_{m=1}^M K_m$-matrix of the inner products between the basis functions. In addition, let $\bm{\Psi}^{1/2}$ denote the square root of $\Psi$. Then, following by \cite{Aguilera2010, Aguilera2016}, it can be shown that the FPLS regression of $Y$ on $\bm{\X}(t)$ (i.e., Model~\eqref{eq:msof}) is equivalent to the PLS regression of $Y$ on $\bm{A} = \bm{D} ( \bm{\Psi}^{1/2} )^\top$ so that at each step of the PLS algorithm, both models produce the same PLS components. 

Let $\widetilde{\bm{\xi}}^{(h)}$ denote the matrix of PLS components (obtained from the regression problem of $Y$ on $\bm{A}$) corresponding to the matrix of eigenvectors $\bm{W}^{(h)}$ obtained at step $h$. Then, Model~\eqref{eq:msof} can be approximated by the following linear regression model:
\begin{equation}\label{eq:appsof}
Y = \bm{1} \gamma_0 + \widetilde{\bm{\xi}}^{(h)} \bm{\gamma} = \bm{1} \gamma_0 + \bm{D} \bm{\Psi} \bm{\beta},
\end{equation}
where $\bm{1}$ is the unit matrix, $\gamma_0$ is the intercept, $\bm{\gamma}$ is the matrix of regression coefficients of $Y$ on $\widetilde{\bm{\xi}}^{(h)}$, and $\bm{\beta} = ( \Psi^{-1/2} )^\top \bm{W}^{(h)} \bm{\gamma}$ is the basis expansion coefficients of the parameter function. Let $\widehat{\bm{\beta}}^{(h)} = ( \Psi^{-1/2} )^\top \bm{W}^{(h)} \widehat{\bm{\gamma}}$, where $\widehat{\bm{\gamma}}$ is the LS estimate of $\bm{\gamma}$. Finally, the parameter function in Model~\eqref{eq:msof} can be approximated as follows:
\begin{equation}\label{eq:bhat}
\widehat{\bm{\beta}}^{(h)}(t) = \bm{\psi}(t) \widehat{\bm{\beta}}^{(h)}.
\end{equation}

\section{The RFPLS method}\label{sec:rfpls}

In the FPLS method summarized in Section~\ref{sec:methodology}, the eigenfunctions $\bm{w}(t)$ defining the FPLS components $\bm{\xi}$ are computed via the usual covariance operator optimized by the LS loss function. In addition, the extracted PLS eigenvectors computed from the basis expansion coefficients of the functional predictors are also based on the LS estimator. Further, in the last steps of FPLS/PLS algorithms, the parameters of the regression model $Y$ on the extracted PLS components are estimated via the LS estimator. However, the LS may be sensitive to outlying observations. In the presence of outliers, the use of the LS estimator leads to a biased estimate of the functional regression coefficients and produces poor predictions of the response variable. This paper proposes a robust PLS method that uses the partial robust M-regression algorithm and the M-estimator along with Tukey's bisquare loss function to robustly obtain the FPLS components and estimate the parameter function, respectively.

Let us now consider the PLS regression of $Y$ on the random vector $\bm{A} = \bm{D} ( \bm{\Psi}^{1/2} )^\top$, which provides an approximate model to the scalar-on-function regression model in~\eqref{eq:msof}, i.e., the linear regression model in~\eqref{eq:appsof}:
\begin{equation*}
Y = \bm{1} \gamma_0 + \widetilde{\bm{\xi}}^{(h)} \bm{\gamma}.
\end{equation*}
For this model, a robust estimate of $\bm{\gamma}$ can easily be obtained using M-estimator. However, in this case, the estimated basis expansion coefficients of the parameter function $\widehat{\bm{\beta}} = ( \Psi^{-1/2} )^\top \bm{W}^{(h)} \widehat{\bm{\gamma}}$ will only be robust to vertical outliers. To obtain an estimate for $\bm{\beta}$ that is robust to both vertical outliers and leverage points, the eigenvectors and PLS components should also be robustly determined. For this purpose, we consider the partial robust M-regression method of \cite{Serneels2005} which down-weights the effects of outliers in the predictor space by obtaining the PLS components via a robust covariance operator, i.e., $\text{Cov}_{r}(Y, \bm{A})$. In this method, the PLS components extracted in each step are adjusted by a weight vector, whose values are determined based on the residuals computed in each step. 

Let $\bm{\xi}_{r}^{(h)}$ denote the matrix of FPLS components computed by partial robust M-regression corresponding to the weights' matrix $\bm{W}_{r}^{(h)}$ obtained at step $h$. Then, Model~\eqref{eq:msof} can be approximated by the robustly obtained PLS components via the following linear regression model:
\begin{equation}\label{eq:robappsof}
Y = \bm{1} \delta_{0} + \bm{\xi}_{r}^{(h)} \bm{\delta} = \bm{1} \delta_{0} + \bm{D} \bm{\Psi} \bm{\beta}.
\end{equation}
For the final estimate of $\bm{\delta}$, we consider the M-estimator along with Tukey's bisquare loss function $\rho(\cdot)$, i.e.,
\begin{equation*}
\widehat{\bm{\delta}}_{\rho} = \underset{\begin{subarray}{c}
  \bm{\delta}	
  \end{subarray}}{\argmin} \sum_{i=1}^n \rho (Y_i - \bm{\xi}_{ri}^{(h)} \bm{\delta} ),
\end{equation*}
where the bisquare loss function is optimized using the tuning parameter selection algorithm proposed by \cite{Wang2007}. Then, the robust basis expansion coefficients of the regression coefficient function $\bm{\beta}(t)$ is obtained by
\begin{equation*}
\widehat{\bm{\beta}}_{r}^{(h)} = ( \Psi^{-1/2} )^\top \bm{W}_{r}^{(h)} \widehat{\bm{\delta}}_{\rho}.
\end{equation*}
Finally, the robust estimate of the parameter functions $\widehat{\bm{\beta}}_{r}^{(h)}(t) = [ \widehat{\bm{\beta}}_{r 1}^{(h)}(t), \ldots, \widehat{\bm{\beta}}_{r M}^{(h)}(t) ]^\top$ are obtained as follows:
\begin{equation}\label{eq:robbhat}
\widehat{\bm{\beta}}_{r}^{(h)}(t) = \bm{\psi}(t) \widehat{\bm{\beta}}_{r}^{(h)}.
\end{equation}
Comparing the estimated parameter functions in~\eqref{eq:bhat} and~\eqref{eq:robbhat}, the latter one is robust to both vertical outliers and leverage points since $\widehat{\bm{\beta}}_{r}^{(h)}$ is computed based on $\bm{W}_{r}^{(h)}$ and $\widehat{\bm{\delta}}_{\rho}$ which are obtained via M-estimator. On the other hand, since the former is computed based on the LS estimator, it is neither robust to vertical outliers nor leverage points. Similarly, compared with the approximate model in~\eqref{eq:appsof}, the approximate model obtained by the M-estimator in~\eqref{eq:robappsof} produces robust predictions for the scalar response $Y$.

The performance of the RFPLS is affected by the number of RFPLS components $h$ used to estimate Model~\eqref{eq:msof}. Thus, the optimum number $h$ should be determined based on a data-driven criterion, such as prediction accuracy. To this end, we consider the following $k$-fold cross-validation procedure. First, divide the data into $k = 1, 2, \ldots, N$ subgroups with roughly the same sample sizes. Using each subgroup as the test sample and the remaining $N-1$ subgroups as the training sample, we implement the RFPLS method on the training sample based on $h = 1, 2, \ldots, H$ number of RFPLS components. Finally, for each of the $H$ models, predict the response variable in the test sample and calculate the $k$-fold trimmed MSPE as follows:
\begin{equation*}
\text{MSPE} = \frac{1}{n^*} \sum_{k=1}^N \sum_{i=1}^{n_k^*} (Y_i^{(k)} - \widehat{Y}_{i}^{(k)} )^2,
\end{equation*}
where $\left\lbrace n_k^* \subset \left\lbrace 1, \ldots, n_k \right\rbrace, ~\vert n_k^* \vert = [ \alpha n_k ] \right\rbrace$ ($\alpha = 0.1$), $n_k = \lfloor n/N \rfloor$ denotes an integer value that rounds down $n/N$, , $n^* = \sum_{k=1}^N n_k^*$, and $Y_i^{(k)}$, and $\widehat{Y}_{i}^{(k)}$ respectively denote the observed and predicted (calculated based on $h$ RFPLS components) response for $i\textsuperscript{th}$ individual in subgroup $k$. The optimum number of RFPLS components is then determined based on the minimum MSPE value. The technical details of the scalar-on-multiple-function PLS regression and the proposed method are presented in Appendix~\ref{sec:app_A}.

\section{Numerical results} \label{sec:results}

We perform several Monte Carlo experiments under different data generation processes and two chemometric data analyses:
\begin{inparaenum}
\item[1)] glucose concentration spectrometric data \citep[available in the \texttt{R} package ``chemometrics''][]{chem} and
\item[2)] sugar process data (available at \url{www.models.life.ku.dk/sugar_process})
\end{inparaenum}
to evaluate the empirical performance of the RFPLS method. We aim to compare the estimation and predictive performance of our RFPLS method with FPLS and FPC, as well as classical PLS of \cite{simpls} (SIMPLS) and the robust PLS approach of \cite{AA17} (RSIMPLS), with the experiments and data analyses. In our numerical analyses, the results for the SIMPLS method are obtained using the \texttt{R} package ``pls'' \citep{plsp} and the results for the RSIMPLS method are obtained using the \texttt{R} code presented by \cite{AA17}. For the FPC regression, the model is built based on the FPC scores obtained from the functional predictors \citep[see, e.g.,][for more information]{ramsay1991}. An example \Rlogo\ code for the functional methods considered in this study can be found at \url{https://github.com/UfukBeyaztas/RFPLS}.

\subsection{Monte Carlo experiments}

Throughout the Monte Carlo experiments, we consider two cases for generating the data. In the first case, the data are generated under a smooth data generation process where the data points on the scalar response and functional predictors follow a similar structure. In this case (Case-I), the RFPLS is expected to provide similar performance to its competitors (which will confirm the proposed method's correctness). In the second case, the scalar response and functional predictors' values are contaminated by deliberately inserted outliers at $[1\%, ~5\%, ~10\%]$ contamination levels. In this case (Case-II), the outliers originate from the error terms and functional predictors (vertical outliers ``plus'' leverage points). First, the leverage points are generated by modifying the data generation process considered under Case-I, and the vertical outliers are generated by contaminating the error values. In these cases, the proposed robust method is expected to produce improved finite-sample performance over its competitors.

\begin{figure}[!htb]
  \centering
  \includegraphics[width= 9cm]{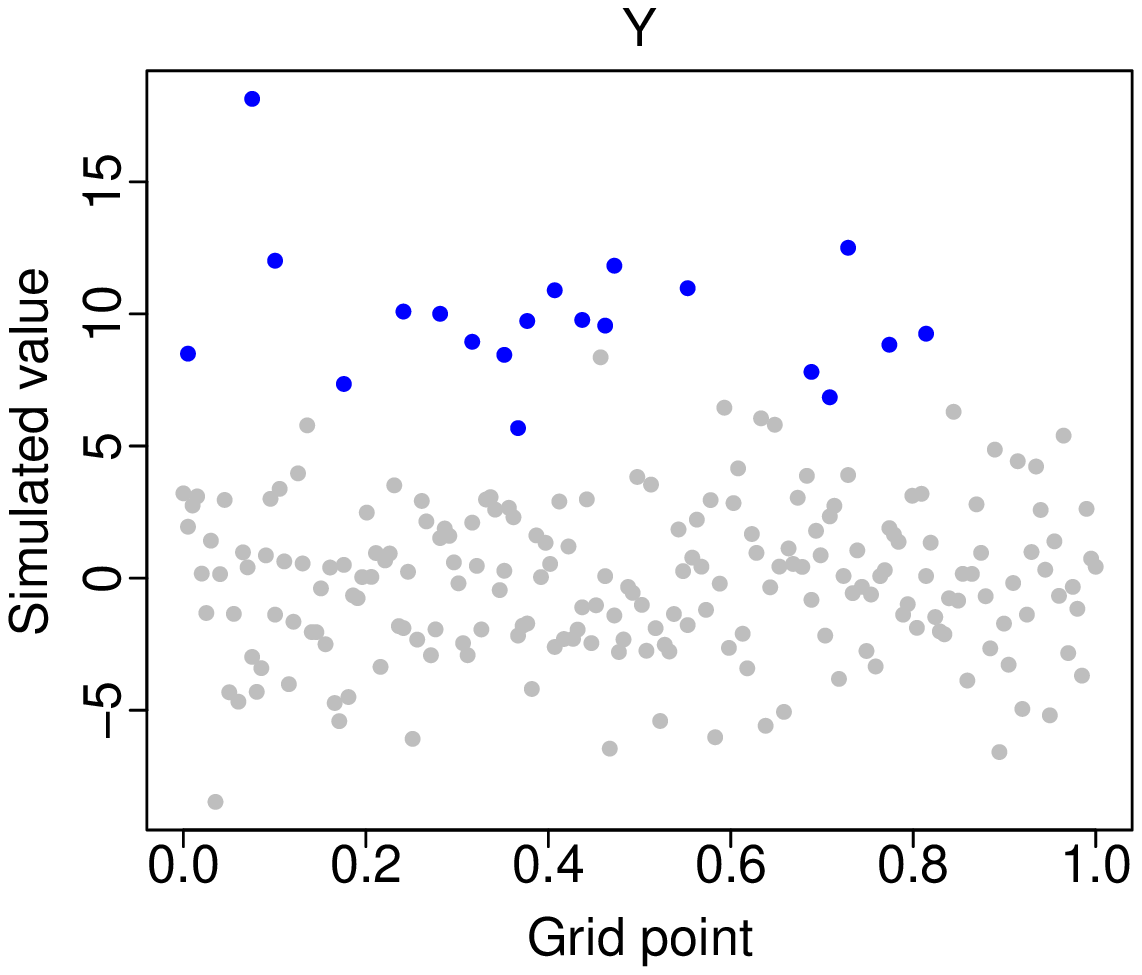}
  \includegraphics[width= 9cm]{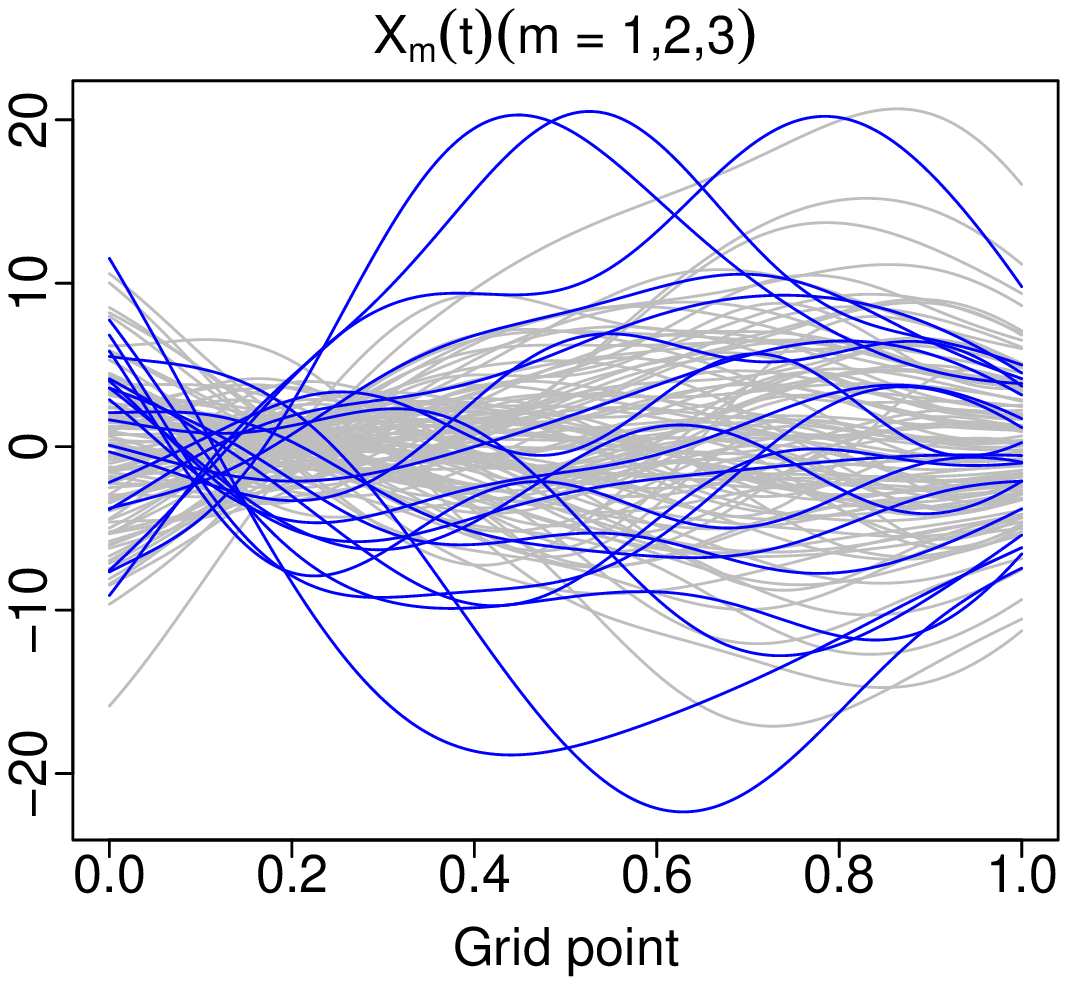}
  \caption{\small{Plots of the generated scalar response and functional variables: observations generated under Case-I (gray) and Case-II (blue).}}\label{fig:Fig_1}
\end{figure}

In the Monte Carlo experiments, we consider a vector of functional predictors consisting of $M = 3$ predictors $\bm{\X}(t) = [ \X_1(t), \X_2(t), \X_3(t) ]^\top$, where each functional predictor consists of $n = 400$ trajectories generated at 200 equally spaced point in the interval $[0,1]$. We consider the following process to generate the trajectories of the functional predictors:
\begin{equation*}
\X_m(t) = \sum_{j=1}^5 \kappa_j \upsilon_j(t),
\end{equation*}
where $\kappa_j \sim \mathcal{N}(0, 4 j^{-3/2})$ and $\upsilon_j(t) = \sin(j \pi t) - \cos(j \pi t)$. In addition, the following processes are considered to generate the parameter functions:
\begin{align*}
\beta_1(t) &= \sin(2 \pi t), \\
\beta_2(t) &= \sin(3 \pi t), \\
\beta_3(t) &= \cos(2 \pi t).
\end{align*}
Then, the values of the scalar response variable are generated as follows:
\begin{equation*}
Y = \sum_{m = 1}^3 \int_0^1 \X_m(t) \beta_m(t) dt + \epsilon,
\end{equation*}
where $\epsilon \sim \mathcal{N}(0,1)$. The generated data are contaminated at $[1\%, ~5\%, ~10\%]$ contamination levels to investigate the robust nature of the proposed method. To generate outliers, first, the randomly selected $n \times [1\%, ~5\%, ~10\%]$ of $\X_m(t)$, for $m = 1,2,3$, are contaminated by the random functions $\widetilde{\X}_m(t)$, where $\widetilde{\X}_m(t)$ are generated similarly to $\X_m(t)$ but using $\upsilon_j(t) = 2 \sin(j \pi t) - \cos(j \pi t)$. Then, the contaminated response values are obtained as follows:
\begin{equation*}
\widetilde{Y} = \sum_{m = 1}^3 \int_0^1 \widetilde{\X}_m(t) \beta_m(t) dt + \widetilde{\epsilon},
\end{equation*}
where $\widetilde{\epsilon} \sim \mathcal{N}(0,10)$. A graphical display of the generated data is presented in Figure~\ref{fig:Fig_1}. 

\begin{figure}[!htb]
  \centering
  \includegraphics[width= 9cm]{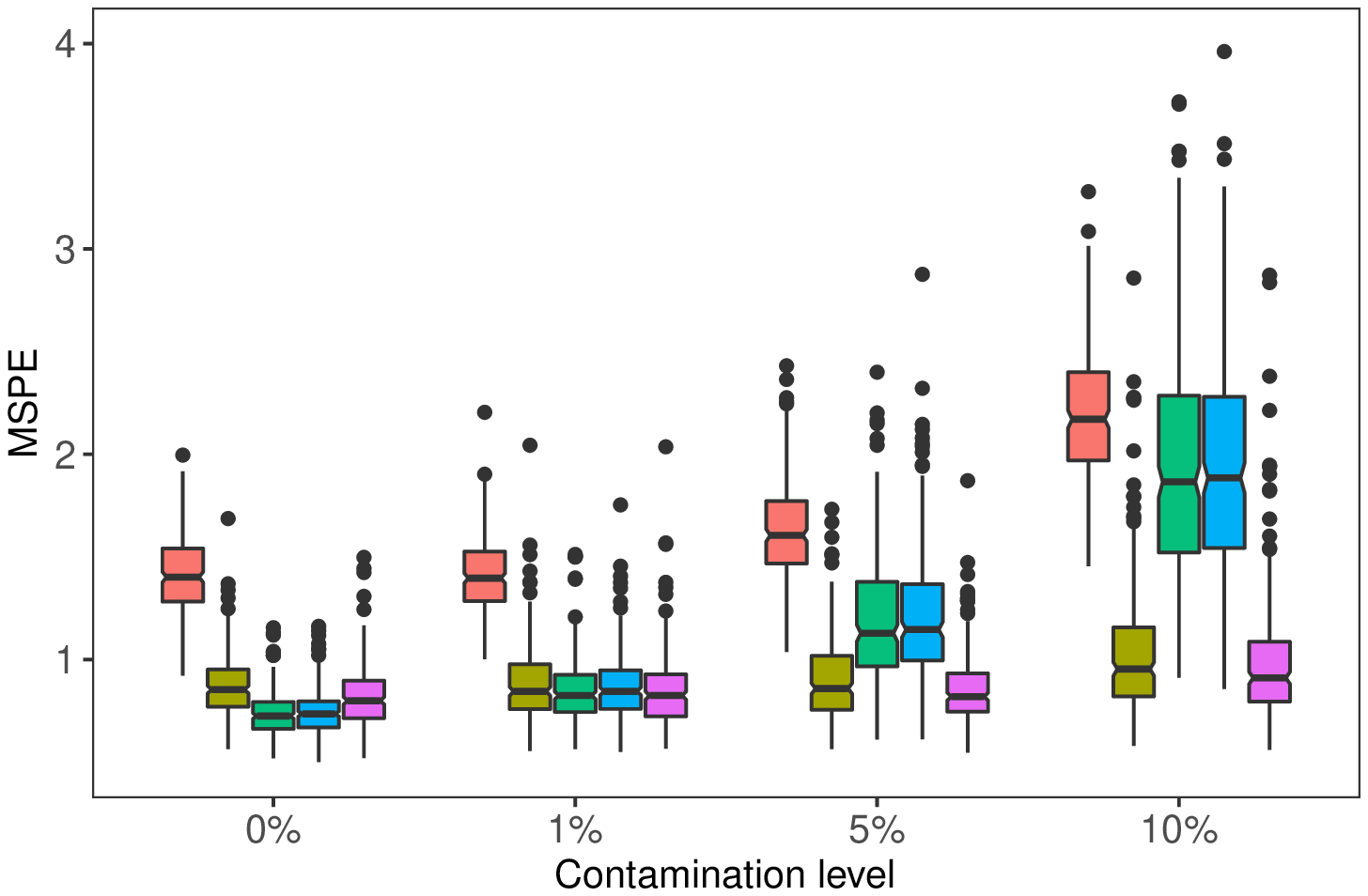}
  \includegraphics[width= 9cm]{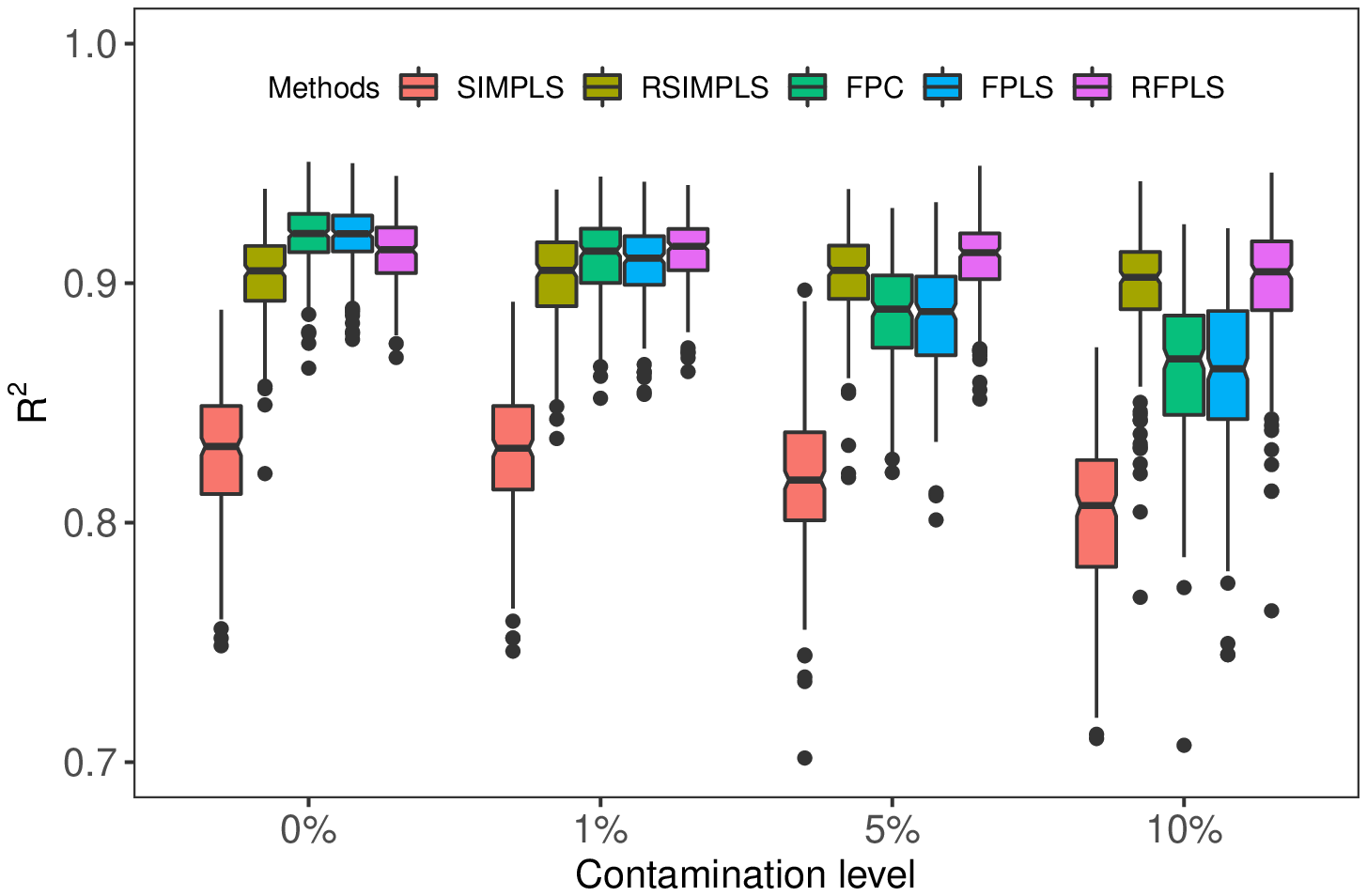}
  \caption{\small{Boxplots of the computed MSPE (left panel) and $R^2$ (right panel) values under Case-I (0\% contamination level in each plot) and Case-II for the multivariate SIMPLS and RSIMPLS and FPC, FPLS, and RFPLS methods}.}
  \label{fig:Fig_2}
\end{figure}

For each case, the following procedure is repeated 1,000 times to compare the finite-sample performance of the functional and multivariate methods. First, the entire generated data are divided into a training and test samples with sizes $n_{\text{train}} = 200$ and $n_{\text{test}} = 200$. The models are constructed to predict the response variable's values. For each replication, the following trimmed MSPE and the trimmed coefficient of determination $R^2$ are computed to measure the out-of-sample predictive performance and relative predictive performance of the methods, respectively:
\begin{equation}
\text{MSPE} = \frac{1}{n_{\text{test}}^*} \sum_{i=1}^{n_{\text{test}}^*} ( Y_i - \widehat{Y}_i )^2, \qquad R^2 = 1 - \frac{1}{n_{\text{test}}^*} \sum_{i=1}^{n_{\text{test}}^*} \frac{(Y_i - \widehat{Y}_i )^2}{( Y_i - \bar{Y}^* )^2},\label{eq:criterion}
\end{equation}
where $\left\lbrace n_{\text{test}}^* \subset \left\lbrace 1, \ldots, n_{\text{test}} \right\rbrace, ~\vert n_{\text{test}}^* \vert = [ \alpha n_{\text{test}} ] \right\rbrace$ ($\alpha = 0.1$), $\widehat{Y}_i$ is the predicted response for the $i^\textsuperscript{th}$ individual in the test sample, and $\bar{Y}^*$ is the trimmed mean. In addition, in each replication, the following relative integrated squared estimation error (RISEE) with respect to the estimated parameter functions is calculated to compare the estimation performance of the functional methods:
\begin{equation*}
\text{RISEE}_m = \frac{\Vert \beta_m(t) - \widehat{\beta}_m(t) \Vert_2^2}{\Vert \beta_m(t) \Vert_2^2},
\end{equation*}
where $\Vert \cdot \Vert_2^2$ denotes $\mathcal{L}_2$ norm, which is approximated by the Riemann sum. We note that in our numerical studies (for both Monte Carlo experiments and empirical data analyses), for the functional methods, the number of $B$-spline basis expansion functions is chosen as 20 to estimate the considered scalar-on-multiple-function regression. Our results are shown in Figure~\ref{fig:Fig_2} and Table~\ref{tab:tab_1}.

From Figure~\ref{fig:Fig_2}, the multivariate SIMPLS of \cite{simpls} produces the worse performance among others. When outliers are not present in the data (i.e., when the contamination level is 0\%), the non-robust functional methods produce slightly better out-of-sample predictive performance (MSPE and $R^2$) than the RSIMPLS and proposed RFPLS methods. On the other hand, when outliers are present in the data, the RSIMPLS and our RFPLS methods produce significantly better out-of-sample predictive performance than the FPLS and FPC. The superiority of the RSIMPLS and our RFPLS over the SIMPLS and classical FPC and FPLS become more prominent as the contamination level increases. From Figure~\ref{fig:Fig_2}, it is evident that both the classical and proposed robust functional methods produce improved performance over their multivariate counterparts. The classical FPC and FPLS significantly outperform multivariate SIMPLS, and the proposed RFPLS slightly outperforms RSIMPLS. While the classical FPC and FPLS produce better results than the multivariate RSIMPLS when no outlier is present in the data, the RSIMPLS produces significantly better results than the non-robust classical functional methods. This is not surprising when considering the robust and non-robust natures of the RSIMPLS and the classical FPC and FPLS methods, respectively.

\begin{table}[!htb]
\begin{center}
\tabcolsep 0.225in
\caption{\small{Computed median RISEE values for FPC, FPLS, and RFPLS method. CL denotes the contamination level.}}\label{tab:tab_1}
\begin{tabular}{@{}cccccccc@{}} 
\toprule
{CL} & {Parameter} & \multicolumn{6}{c}{Method} \\
\midrule
& & \multicolumn{3}{c}{Case-I} & \multicolumn{3}{c}{Case-II} \\
\cmidrule(l){3-5} \cmidrule(l){6-8}
& & FPC & FPLS & RFPLS & FPC & FPLS & RFPLS \\
\midrule
\multirow{3}{*}{0\%} & $\beta_1(t)$ & \textbf{0.132} & 0.166 & 0.201  \\
& $\beta_2(t)$ & \textbf{0.133} & 0.171 & 0.213  \\
& $\beta_3(t)$ & \textbf{0.122} & 0.166 & 0.145  \\
\midrule
\multirow{3}{*}{1\%} & $\beta_1(t)$ & & & & 0.266 & 0.801 & \textbf{0.259}   \\
& $\beta_2(t)$ & & & & 0.290 & 0.563 & \textbf{0.265}   \\
& $\beta_3(t)$ & & & & 0.245 & 0.550 & \textbf{0.152}  \\
\midrule
\multirow{3}{*}{5\%} & $\beta_1(t)$ & & & & 0.342 & 1.129 & \textbf{0.279}   \\
& $\beta_2(t)$ & & & & 0.348 & 1.189 & \textbf{0.302}   \\
& $\beta_3(t)$ & & & & 0.356 & 1.360 & \textbf{0.219}   \\
\midrule
\multirow{3}{*}{10\%} & $\beta_1(t)$ & & & & 0.891 & 1.089 & \textbf{0.515}   \\
& $\beta_2(t)$ & & & & 0.799 & 0.912 & \textbf{0.456}  \\
& $\beta_3(t)$ & & & & 0.685 & 1.075 & \textbf{0.373}  \\
\bottomrule
\end{tabular}
\end{center}
\end{table}

The RISEE values computed for each parameter function under both cases are given in Table~\ref{tab:tab_1}. From Table~\ref{tab:tab_1}, the FPC and FPLS produce better parameter estimates than the proposed method when outliers are not present in the data. On the other hand, the proposed method produces considerably smaller RISEE values than the classical FPC and FPLS, especially when the contamination level is high. From Table~\ref{tab:tab_1}, FPC tends to produce better RISEE values than the FPLS. 

\subsection{Empirical data analyses}

\subsubsection{Glucose concentration spectrometric data}

\begin{figure}[!htb]
  \centering
  \includegraphics[width= 9cm]{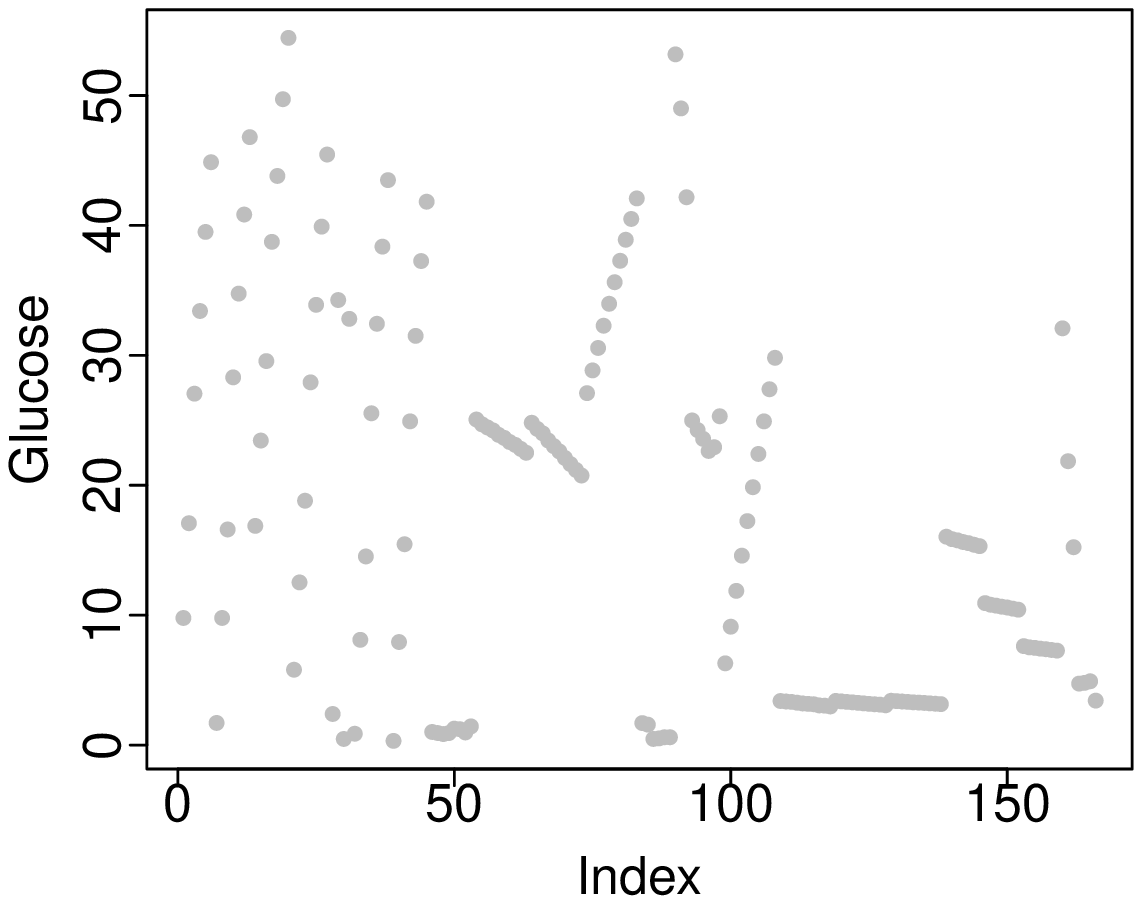}
  \includegraphics[width= 9cm]{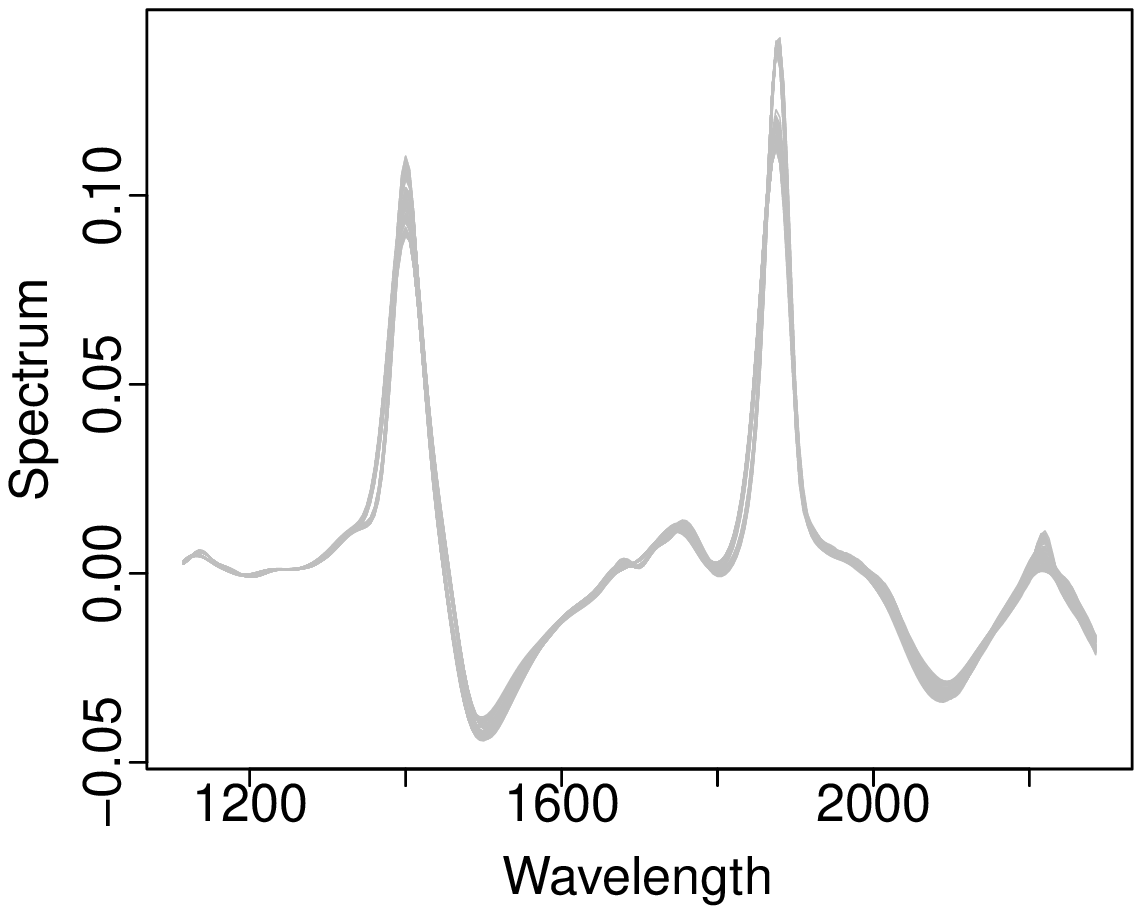}
  \caption{\small{A graphical display of the 166 Glucose concentrations (g/L) (left panel) and first-derivative spectra of laboratory samples (right panel)}.}
  \label{fig:Fig_3}
\end{figure}

We consider the glucose concentration spectrometric data, originally reported by \cite{Liebmann2009}. Two essential compounds in the bioethanol process are glucose, the nutrient for yeast fermentation, and its fermentation product ethanol \citep[see, e.g.,][]{Liebmann2009}. In the bioethanol fermentation experiment, 166 samples were taken during fermentation mashes of different feedstock (rye, wheat, and corn). Near-infrared (NIR) spectra were measured (in 2 nm intervals from 1115-2285 nm, 235 wavelengths in total) from the samples, and the glucose concentrations were computed from the high-performance liquid chromatography. A graphical display of the glucose concentrations and the NIR spectra of 166 mash samples is presented in Figure~\ref{fig:Fig_3}.

It is of great interest to predict glucose concentration for a given NIR spectra curve for this dataset. To evaluate and compare the predictive performance of the methods, the entire dataset is randomly divided into a training sample with size $n_{\text{train}} = 100$ and a test sample with size $n_{\text{test}} = 66$. The models are constructed with the training sample to predict the glucose concentrations in the test sample. This process is repeated 1000 times, and for each replication, the MSPE and $R^2$ are calculated for all the methods. For this dataset, we check the outliers in the scalar response using interquartile range (IQR), i.e., observations below $Q_1 - 1.5 \times IQR$ or above $Q_3 + 1.5 \times IQR$, where $Q_1$ and $Q_3$ are respectively the first and third quartile of the data and $IQR = Q_3 - Q_1$, are called an outlier. Based on the IQR, there is no outlier in the response variable. For the functional predictor, on the other hand, we consider the functional boxplot method proposed by \cite{SunG} available in the \texttt{R} package ``fdaoutlier'' \citep{fdaoutlier}. The functional boxplots results show no outlying curve in the functional predictor. Thus, all the methods are expected to produce a similar performance. We note that for the functional methods, the number of basis expansion functions is selected as $K = 20$ (arbitrarily) in the interval 1115-2285. The functional methods may produce a different performance for different $K$ values, and thus, the optimum number of $K$ may be determined via an information criterion such as generalized cross-validation.

\begin{figure}[!htb]
  \centering
  \includegraphics[width= 9cm]{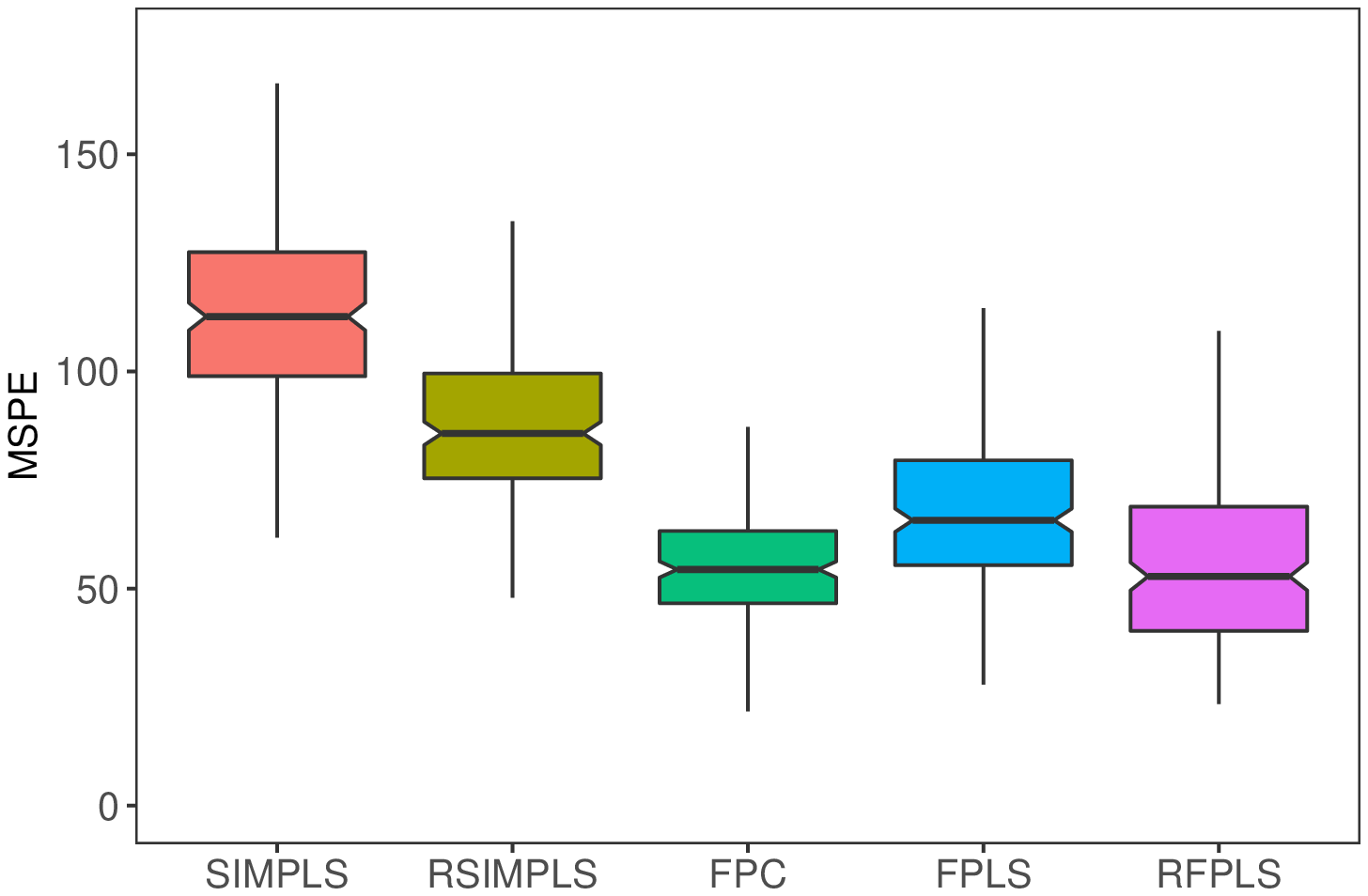}
  \includegraphics[width= 9cm]{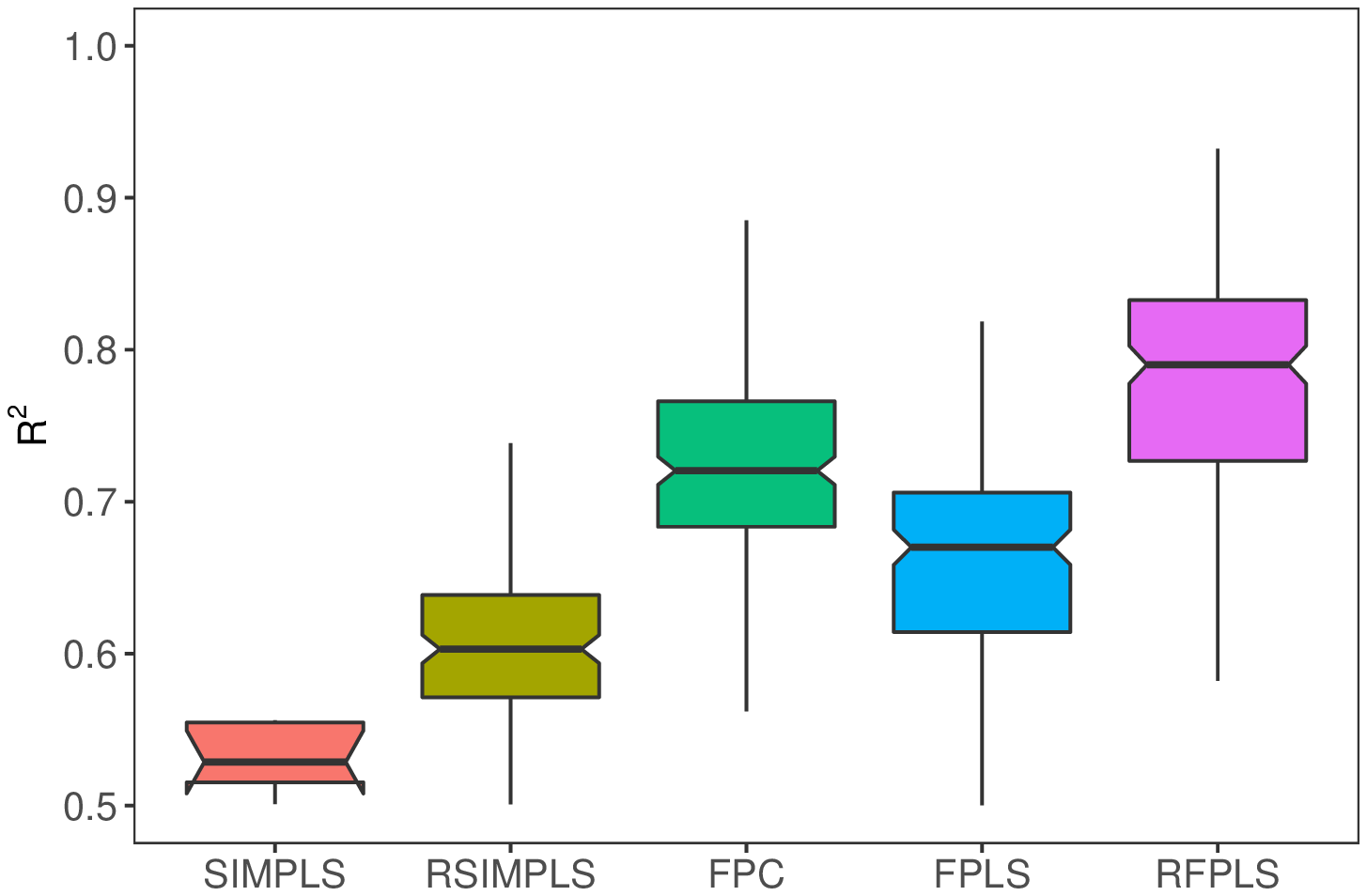}
  \caption{\small{Boxplots of the computed MSPE (left panel) and $R^2$ (right panel) values for the glucose concentration spectrometric data}.}
  \label{fig:Fig_4}
\end{figure}

Our results are presented in Figure~\ref{fig:Fig_4}, which shows that the functional methods outperform multivariate methods so that the functional methods produce roughly 2 and 1.5 times fewer MSPE values compared with multivariate SIMPLS and RSIMPLS, respectively. In addition, the results presented in Figure~\ref{fig:Fig_4} indicate that the proposed method produces slightly better out-of-sample predictive performance than the FPC and FPLS methods. 

\subsubsection{Sugar process data}

\begin{figure}[!htb]
  \centering
  \includegraphics[width=4.5cm]{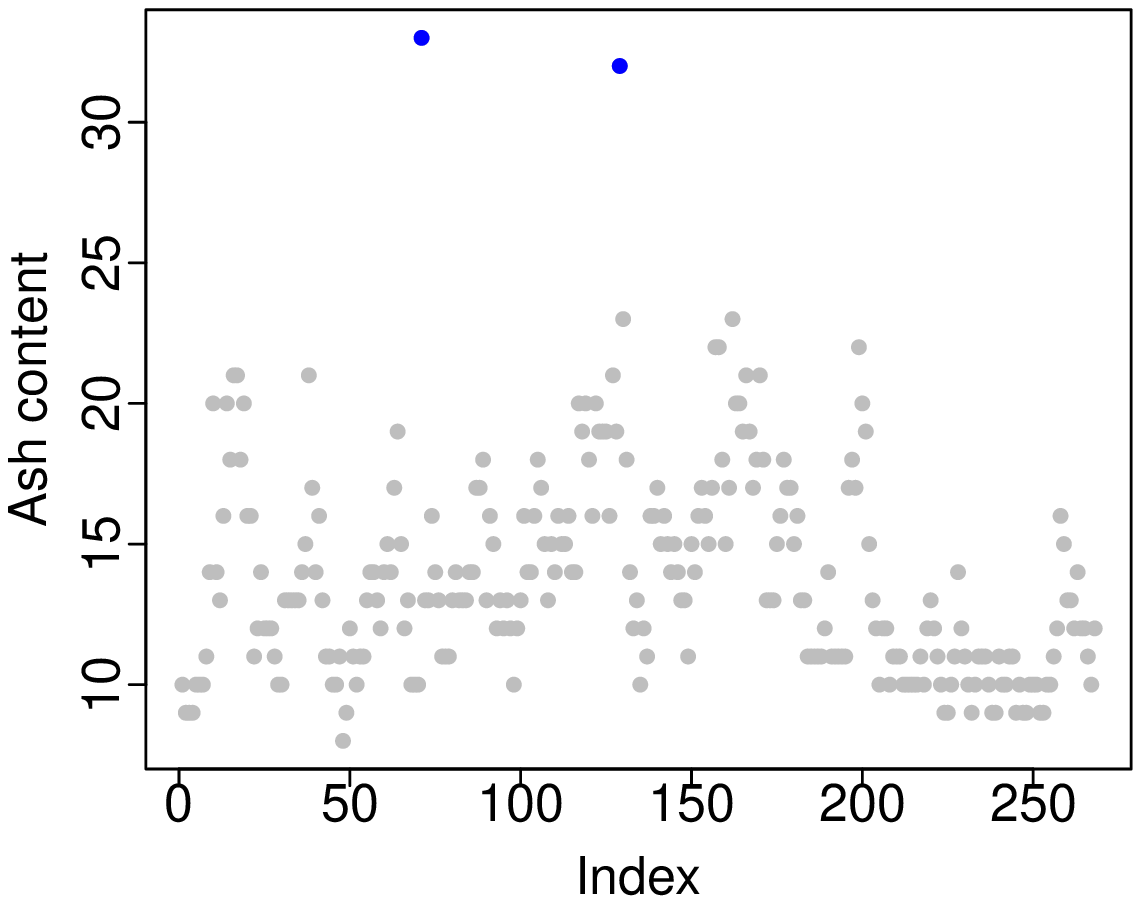}
  \includegraphics[width=4.5cm]{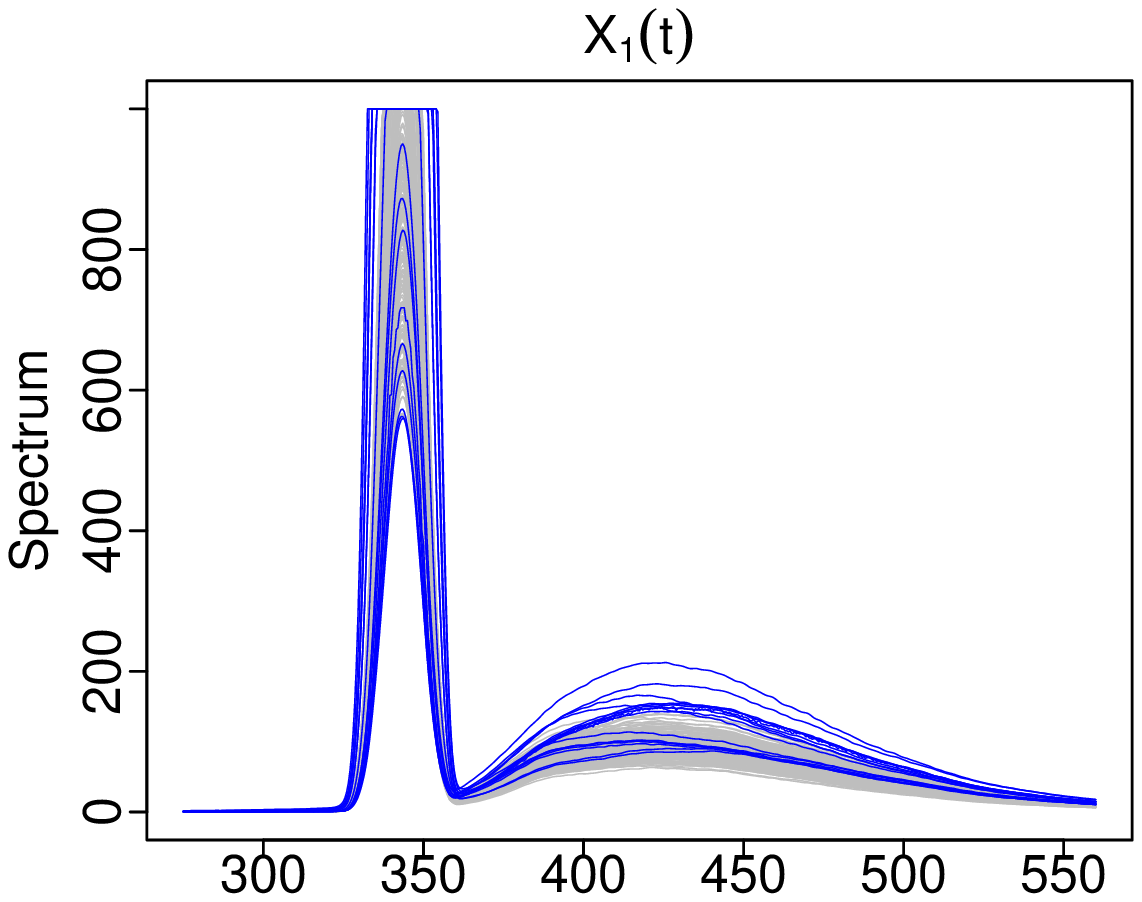}
  \includegraphics[width=4.5cm]{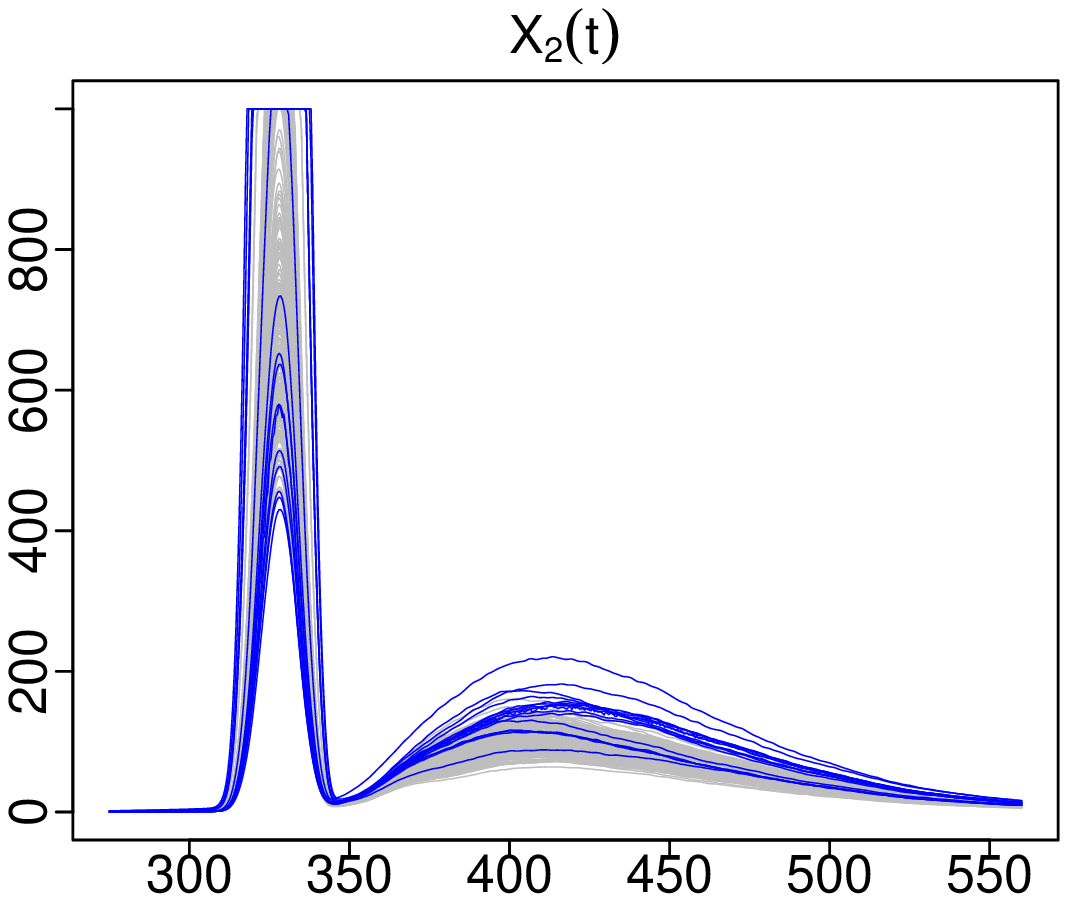}
  \includegraphics[width=4.5cm]{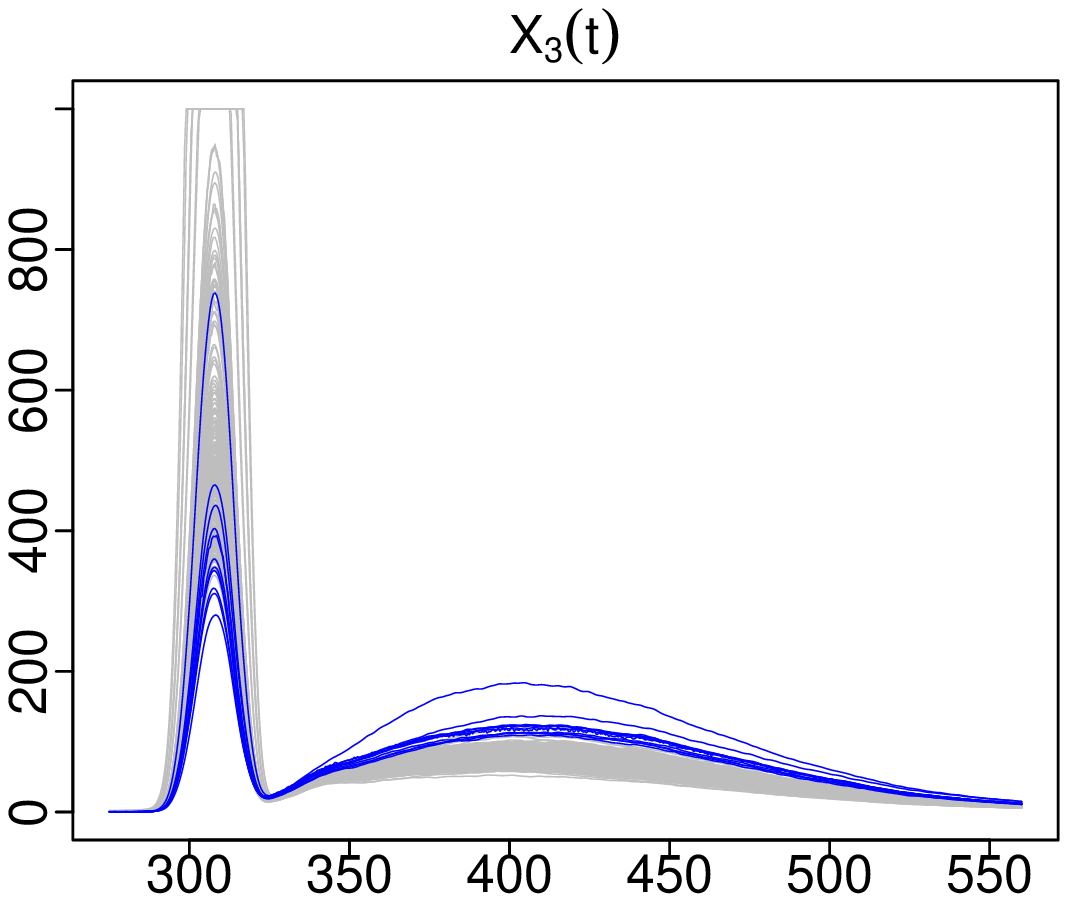}
\\  
  \includegraphics[width=4.5cm]{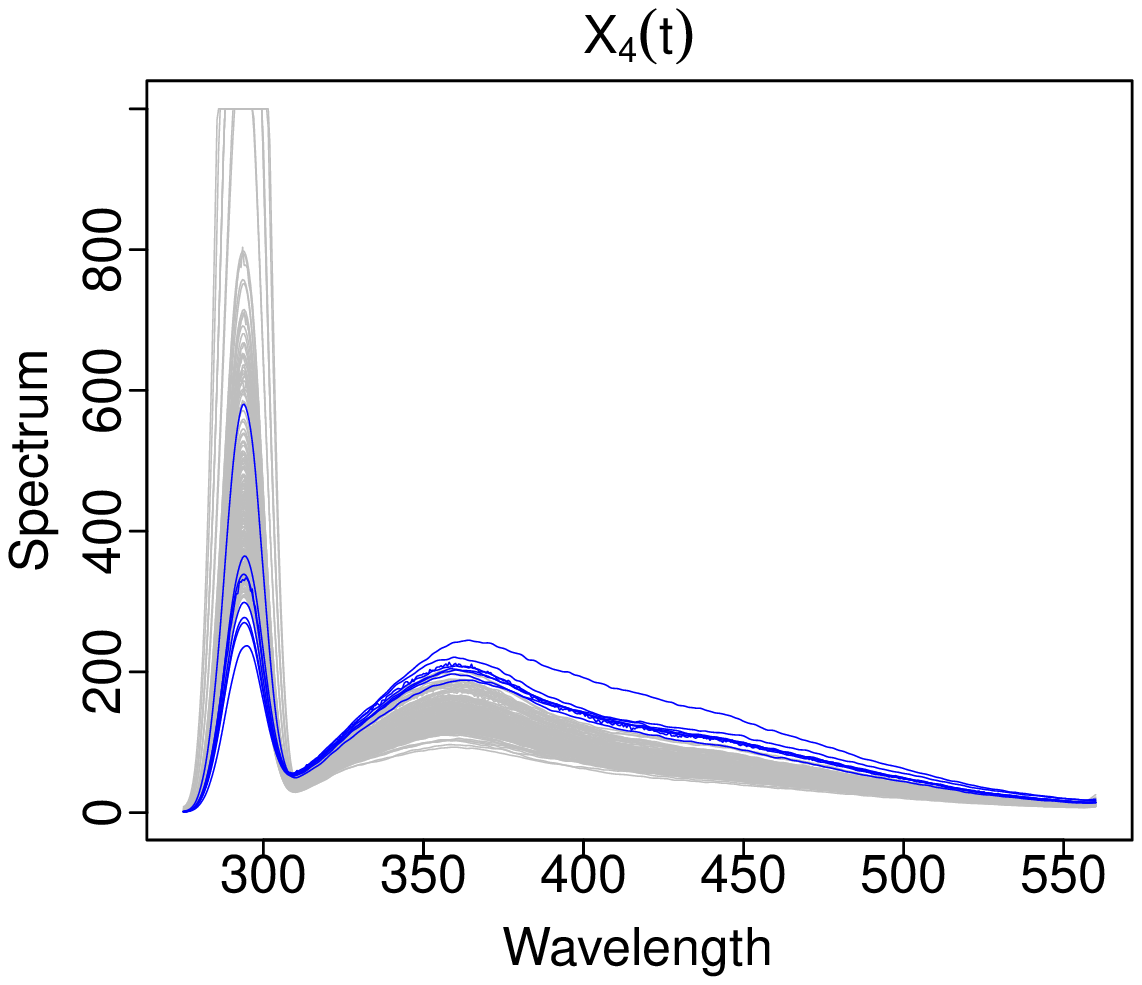}
  \includegraphics[width=4.5cm]{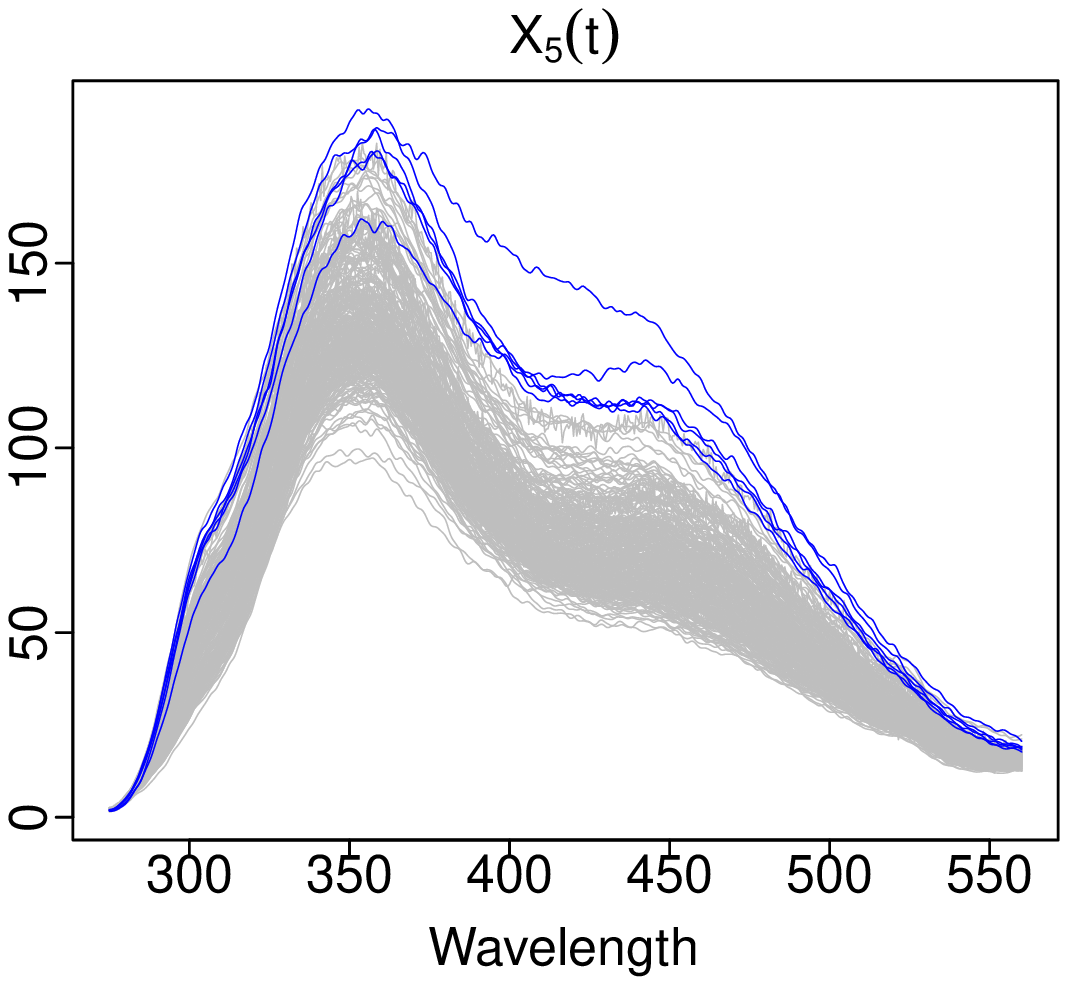}
  \includegraphics[width=4.5cm]{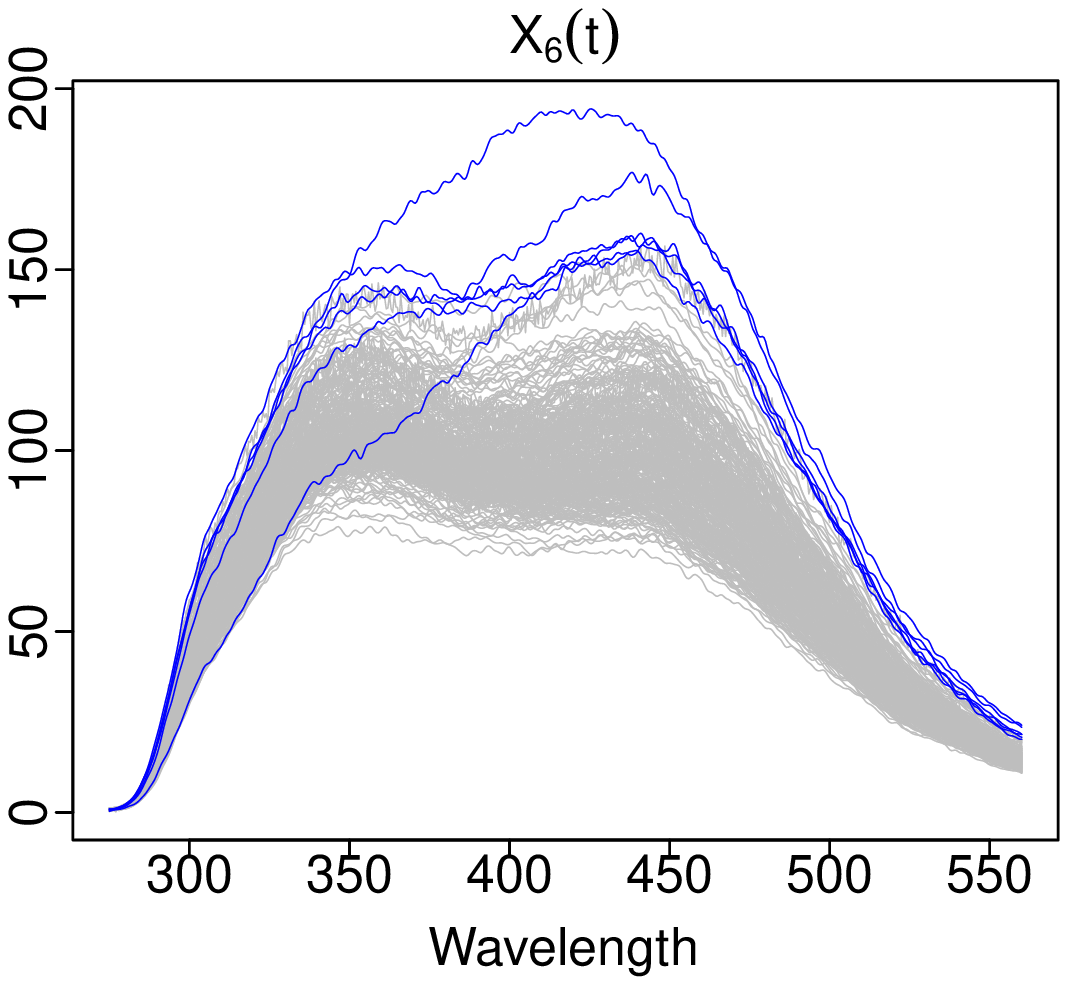}
  \includegraphics[width=4.5cm]{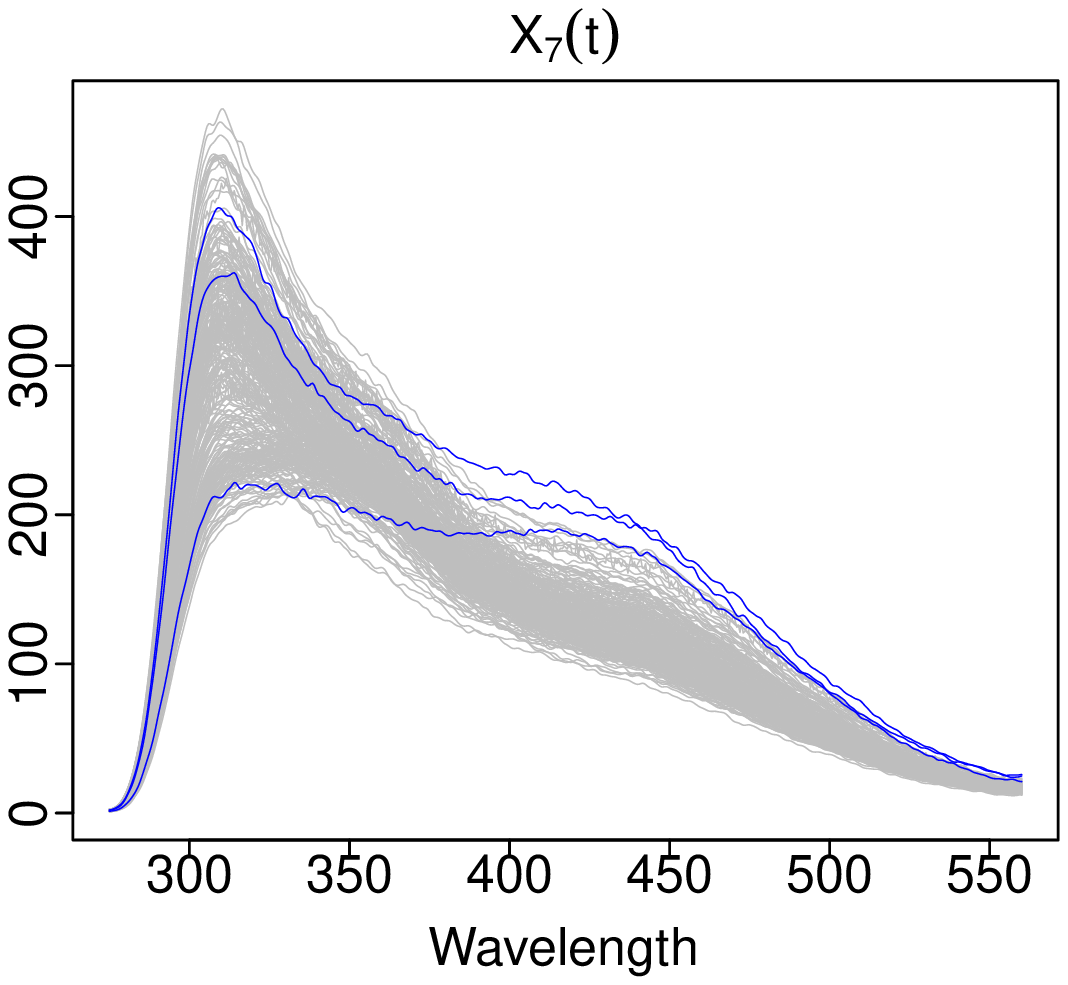}
  \caption{\small{A graphical display of the 268 ash content and the emission spectra of sugar samples obtained at excitation wavelengths $[230, 240, 255, 290, 305, 325, 340]$ $=$ $[\X_1(t), \X_2(t), \X_3(t), \X_4(t), \X_5(t), \X_6(t), \X_7(t)]$. Normal observations are given by gray color while outliers are marked with blue color}.}
  \label{fig:Fig_5}
\end{figure}

We also consider the sugar process data, described initially by \cite{Munck1998} and \cite{Bro1999}. A solution was measured spectro-fluorometrically from the sugar dissolved in un-buffered water in the experiment. For each of the analyzed 268 sugar samples, the emission spectra were measured in 0.5 nm intervals from 275 to 560 (571 wavelengths in total) at seven excitation wavelengths (nm) $[230, 240, 255, 290, 305, 325, 340]$. Also, a quality parameter, ``ash content'', a measure of the number of inorganic impurities in the refined sugar, was measured fluorometrically from the 268 sugar samples. A graphical display of the sugar process data is presented in Figure~\ref{fig:Fig_5}. 

With this dataset, we aim to predict predictions for ash content for given excitation wavelengths \citep[see also][]{Gartheiss2013, Smaga2018}. According to IQR, there are two clear outliers in the response variable; points 32 and 33. In addition, the results of functional boxplots indicate that all the functional predictors include around 4\% - 10\% outliers with small and large magnitudes (refer to Figure~\ref{fig:Fig_5}). These outliers may cause poor predictions for the ash content when the non-robust methods estimate the model. These motivate us to apply our RFPLS method to robustly model ash content and the spectra measured at different excitation wavelengths.

The following procedure is repeated 1,000 times to evaluate and compare the predictive performance of the proposed and classical methods for the sugar process data. First, we randomly divide the entire dataset into a training with size $n_{\text{train}}=100$ and a test sample with size $n_{\text{test}}=168$. With the training sample, we construct the models to predict the ash content in the testing sample. We compute the MSPE and $R^2$ values in~\eqref{eq:criterion} for each replication, from which we compare the methods. For this dataset, for each functional predictor, $K = 20$ B-spline basis expansion functions in the interval 275-560 are used to construct models for the functional methods.

\begin{figure}[!htb]
  \centering
  \includegraphics[width=8.7cm]{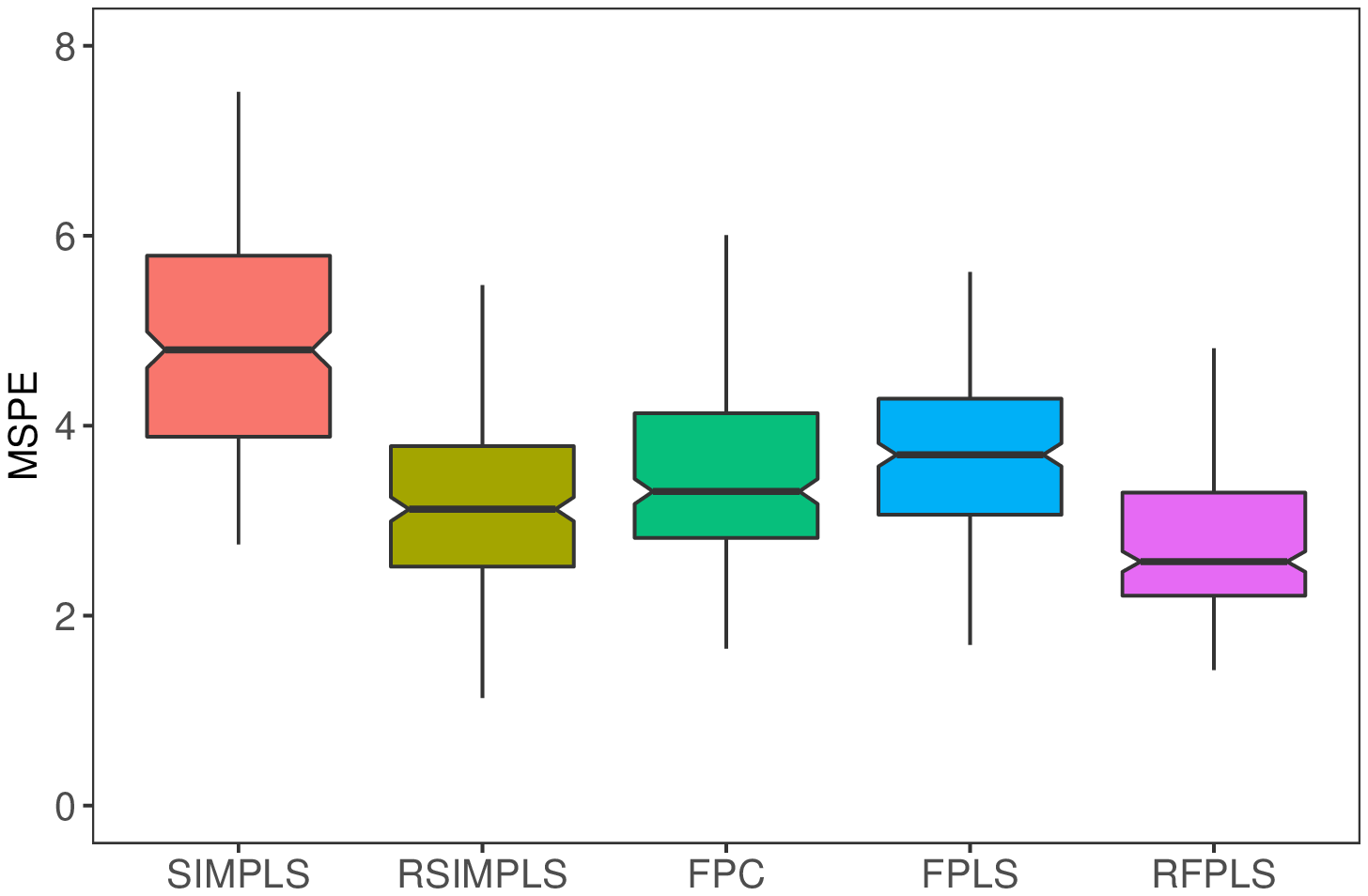}\qquad
  \includegraphics[width=8.7cm]{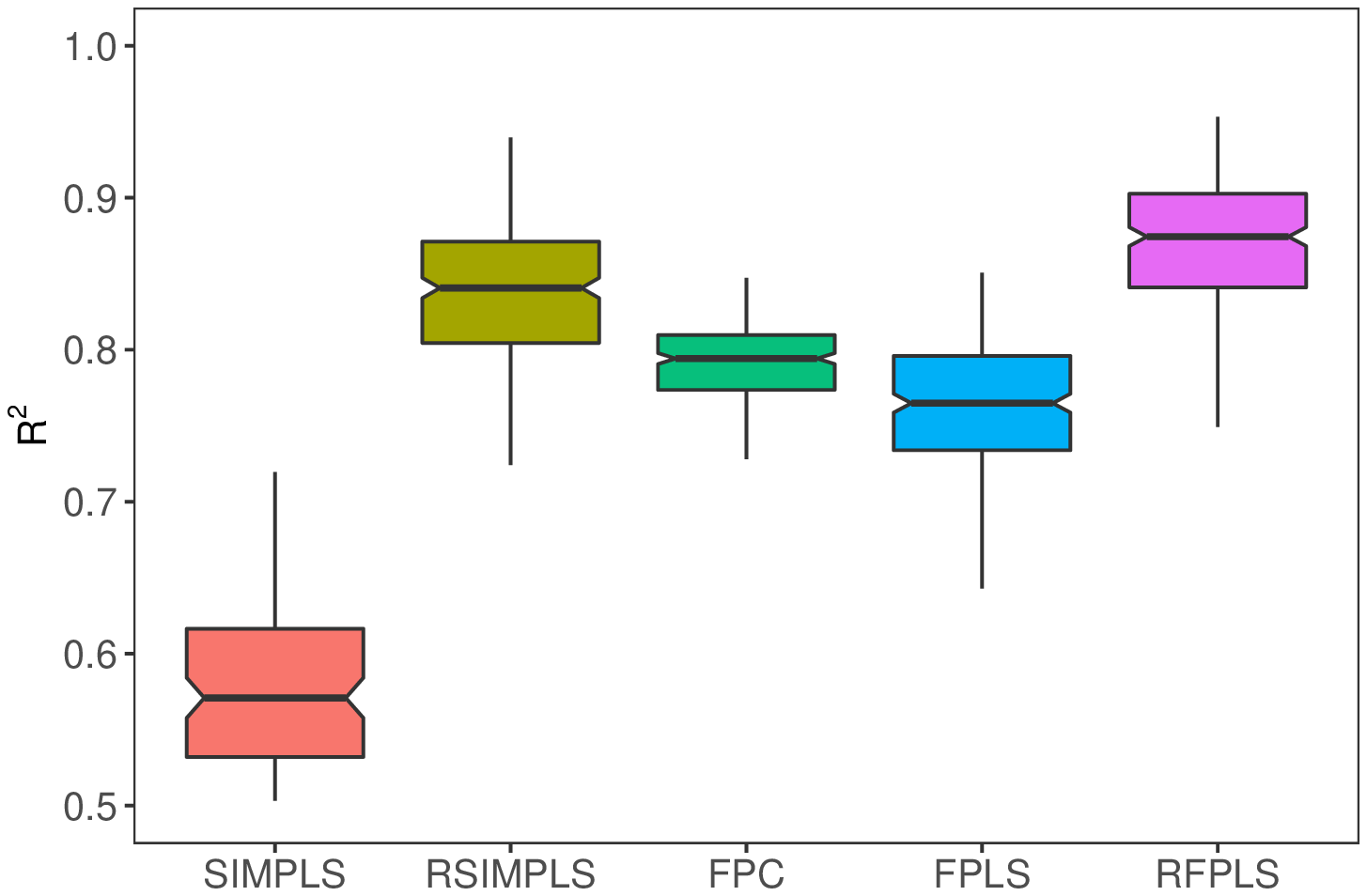}
  \caption{\small{Boxplots of the computed MSPE (left panel) and $R^2$ (right panel) values for the sugar process data}.}
  \label{fig:Fig_6}
\end{figure}

Our results are presented in Figure~\ref{fig:Fig_6}, which indicates that our proposed method produces significantly improved performance than the existing functional and multivariate methods. The results also indicate that the classical functional methods (FPC and FPLS) produce improved performance over the multivariate SIMPLS method. On the other hand, the multivariate RSIMPLS produces better MSPE and $R^2$ values than those of FPC and FPLS because both the response and predictors variables include outliers.

\section{Conclusion} \label{sec:conc}

We propose an RFPLS method that allows more than one functional predictor to estimate the regression coefficients in a scalar-on-function regression model. Our proposed method uses robust M-regression to obtain the FPLS components. Then, the M-estimator and Tukey's bisquare loss function are used to obtain the final estimates. The finite-sample predictive performance of our method is evaluated through several Monte Carlo experiments and two chemometrics datasets. We compare our results with those of classical FPLS and FPC and multivariate SIMPLS and RSIMPLS methods. Our numerical results demonstrate that the proposed method produces similar predictive performance with the classical methods when outliers are not present in the data. Simultaneously, it outperforms the classical methods when outliers contaminate the predictor and response variables.

There are several directions that our proposed method can be further extended. For example:
\begin{inparaenum}
\item[1)] In our proposal, we consider only functional predictors. However, one may prefer a mixed data type consisting of functional and scalar predictors to explain the scalar response variation, such as a partial functional linear model \citep[see, e.g.,][]{Shin09}. Our proposed method can be extended to this model when the scalar predictors are included.
\item[2)] The RFPLS method can also be extended to the function-on-function-regression model. Both the response and predictor variables consist of random curves, as an alternative to \cite{PredSc} and \cite{BS20}.
\item[3)] In the present study, we only consider the main effects of the functional predictors. However, recent studies \citep[see, e.g.,][]{Fuchs, Usset, LuoQi, SunWang, Matsui2020, BS22} have shown that functional regression models with quadratic and interaction effects of the functional predictors perform better than main effects models. When the functional predictors include outliers, the effects of outliers on the model may be greater in the functional predictors' quadratic and interaction terms than the main effects. The proposed method can robustly estimate the functional regression models with quadratic and interaction effects.
\item[4)] To further investigate the predictive performance of our proposed method, an appropriate robust bootstrap method can be used with the RFPLS to construct a prediction interval for the scalar response variable. 
\end{inparaenum}

\section*{Acknowledgments}

We thank Editor, Professor John H. Kalivas, and two reviewers for their constructive comments, which have helped us produce a significantly improved paper. This work was supported by The Scientific and Technological Research Council of Turkey (TUBITAK) (grant no: 120F270).

\newpage
\bibliographystyle{agsm}
\bibliography{rfpls}

\newpage
\appendix

\begin{center}
\large Appendixes
\end{center}
We present the technical details of the FPLS and the proposed RFPLS methods for the scalar-on-multiple-function regression model.

\section{The FPLS for the scalar-on-multiple function regression}\label{sec:app_A}

For the scalar-on-multiple-function regression model~\eqref{eq:msof}, the FPLS components, $\bm{\xi}$, are obtained as the solutions of Tucker's criterion
\begin{align}
&\underset{\begin{subarray}{c}
  \bm{w}(t) \in \mathcal{L}_2^M[0,1] \\ \Vert w_m(t) \Vert = 1,~\forall m = 1, \ldots, M
  \end{subarray}}{\argmax} \text{Cov}^2 \left( Y, \int_0^1 \bm{\X}^\top(t) \bm{w}(t) dt \right), \label{eq:cov} \\
  \Leftrightarrow& \underset{\begin{subarray}{c}
  \bm{w}(t) \in \mathcal{L}_2^M[0,1]
  \end{subarray}}{\argmin}
\text{E}^2 \left[ \varphi \left( Y - \int_0^1 \bm{\X}^\top(t) \bm{w}(t) dt \right) \right], \label{eq:ecov}
\end{align}
where $\varphi(u) = u^2$ is the least-squares loss function. Let us denote the cross-covariance operators $\mathcal{C}_{Y \bm{\X}}$ and $\mathcal{C}_{\bm{\X} Y}$ to evaluate the contribution of $\bm{\X}(t)$ to $Y$ as follows \citep{PredSap, Aguilera2016}:
\begin{align*}
\mathcal{C}_{Y \bm{\X}}: \mathcal{L}_2^M[0,1] \rightarrow \mathbb{R},& \qquad \bm{f} \xrightarrow{\mathcal{C}_{Y \bm{\X}}} x = \int_0^1 \text{Cov} \left( \bm{\X}(t), Y \right) \bm{f}(t) dt, \\
\mathcal{C}_{\bm{\X} Y}: \mathbb{R} \rightarrow \mathcal{L}_2^M[0,1],& \qquad x \xrightarrow{\mathcal{C}_{\bm{\X} Y}} \bm{f}(t) = x~\text{Cov} \left( \bm{\X}(t), Y \right).
\end{align*}
From the $\mathcal{L}_2$ continuity of $\bm{\X}(t)$ (i.e., the $\mathcal{L}_2$ continuity of $\X_m(t),~\forall~m=1,\ldots,M$), $\mathcal{U} = \mathcal{C}_{\bm{\X} Y} \circ \mathcal{C}_{Y \bm{\X}}$ is defined as self-adjoint, positive, and compact operator \citep{PredSap} and its spectral analysis produces a countable set of positive eigenvalues $\lambda$ associated to orthonormal basis of eigenfunctions $\bm{w}$ as a solution of $\mathcal{U} \bm{w} = \lambda \bm{w}$. The optimization problem of~\eqref{eq:cov} or~\eqref{eq:ecov} can be expressed as
\begin{equation*}
\underset{\begin{subarray}{c}
  \bm{w}(t) \in \mathcal{L}_2^M[0,1]
  \end{subarray}}{\max} \frac{\langle \mathcal{U} \bm{w},~\bm{w} \rangle_{\mathcal{H}^M}}{\langle \bm{w},~\bm{w} \rangle_{\mathcal{H}^M}}
\end{equation*}
and, the solution to Tucker's criterion corresponds to the eigenfunction generated by $\mathcal{U}$ associated to its largest eigenvalue, denoted by $\lambda_{\max}$, i.e., $\mathcal{U} \bm{w} = \lambda_{\max} \bm{w}$. In what follows, the FPLS components are obtained as the linear functionals of $\bm{\X}(t)$ and $\bm{w}$, i.e., $\bm{\xi} = \int_0^1 \bm{\X}^\top(t) \bm{w}(t) dt$.

As noted in Section~\ref{sec:methodology}, the FPLS is an iterative method, and at each iteration, the FPLS component is obtained by taking into account the information gathered from the previous iteration. More precisely, let $h = 1, 2, \ldots$ denote the iteration number and let $Y^{(h)}$ and $\bm{\X}^{(h)}(t)$ denote the residuals obtained from the following regression problems:
\begin{align*}
Y^{(h)} &= Y^{(h-1)} - c^{(h-1)} \bm{\xi}^{(h-1)}, \\
\bm{\X}^{(h)}(t) &= \bm{\X}^{(h-1)}(t) - \bm{p}^{(h)}(t) \bm{\xi}^{(h)},
\end{align*}
where $c^{(h)} = \text{E}[Y^{(h-1)} \bm{\xi}^{(h)}] / \text{E}[(\bm{\xi}^{(h)})^2]$, $\bm{p}^{(h)}(t) = \text{E}[\bm{\X}^{(h-1)}(t) \bm{\xi}^{(h)}] / \text{E}[\bm{(\xi}^{(h)})^2]$, and $Y^{(h-1)} = Y$ and $\bm{\X}^{(h-1)}(t) = \bm{\X}(t)$ when $h = 1$. Then, at step $h$, the $h^\textsuperscript{th}$ eigenfunction is obtained as a solution of
\begin{equation*}
\bm{w}^{(h)}(t) = \underset{\begin{subarray}{c}
  \bm{w}(t) \in \mathcal{L}_2^M[0,1] \\ \Vert w_m(t) \Vert = 1,~\forall m = 1, \ldots, M
  \end{subarray}}{\argmax} \text{Cov}^2 \left( Y^{(h-1)}, \int_0^1 [\bm{\X}^{(h-1)}(t)]^{\top} \bm{w}(t) dt \right),
\end{equation*}
and the corresponding FPLS component is computed as
\begin{equation*}
\bm{\xi}^{(h)} = \int_0^1 [\bm{\X}^{(h-1)}(t)]^{\top} \bm{w}^{(h)}(t) dt,
\end{equation*}
i.e., it is the largest eigenfunction corresponding to the largest eigenvalue of $\mathcal{U}_{h-1} = \mathcal{C}_{\bm{\X} Y}^{h-1} \circ \mathcal{C}_{Y \bm{\X}}^{h-1}$, where $\mathcal{C}_{\bm{\X} Y}^{h-1}$ and $\mathcal{C}_{Y \bm{\X}}^{h-1}$ are the cross-covariance operators of $Y^{(h-1)}$ and $\bm{\X}^{(h-1)}$, respectively.

The FPLS eigenfunctions and therefore the FPLS components cannot be computed directly because of the infinite-dimensional nature of the functional predictors. To overcome this problem, we approximate them in a finite-dimensional space of basis expansion coefficients, see e.g., Section~\ref{sec:methodology}. Let us consider the basis expansion form of the functional predictors' vector, $\bm{\X}(t) = \bm{D} \bm{\psi}^\top(t)$. From the spectral analysis of $\mathcal{U}$, i.e., $\mathcal{U} \bm{w} = \lambda \bm{w}$, the FPLS eigenfunctions and the corresponding FPLS components are approximated in the basis $\bm{\psi}(t)$ as $\bm{w}(t) = \bm{\psi}(t) \bm{w}$ and $\bm{\beta}(t) = \bm{\psi}(t) \bm{\beta}$, respectively. In addition, the cross-covariance operators can be expressed in terms of the basis expansion of $\bm{\X}(t)$ as follows:
\begin{align*}
\mathcal{C}_{Y \bm{\X}}: \mathcal{L}_2^M[0,1] \rightarrow \mathbb{R},& \qquad \bm{f} \xrightarrow{\mathcal{C}_{Y \bm{\X}}} x = \bm{\Sigma}^\top_{\bm{D} Y} \bm{\Psi} \bm{f}, \\
\mathcal{C}_{\bm{\X} Y}: \mathbb{R} \rightarrow \mathcal{L}_2^M[0,1],& \qquad x \xrightarrow{\mathcal{C}_{\bm{\X} Y}} \bm{f}(t) = x~ \bm{\Sigma}_{\bm{D} Y},
\end{align*}
where $\bm{\Sigma}_{\bm{D} Y} = \left[ \bm{\Sigma}^{(1)}_{\bm{d}_1 Y}, \ldots, \bm{\Sigma}^{(M)}_{\bm{d}_M Y} \right]^\top$ with $\bm{\Sigma}^{(m)} = ( \sigma_{mk} )_{K_m \times 1}$ and $\sigma_{mk} = \text{E} [ d_{mk} Y ]$ for $m = 1, \ldots, M$ and $k = 1, \ldots, K_m$ is the cross-covariance matrix between $\bm{D}$ and $Y$ and $\bm{\Psi} = \int_0^1 \bm{\psi}(t) \bm{\psi}^\top(t) dt$ is the symmetric block-diagonal $\sum_{m=1}^M K_m \times \sum_{m=1}^M K_m$-matrix of the inner products between the basis functions. Note that for the orthogonal basis functions such as Fourier basis, $\bm{\Psi} = \mathbb{1}$, i.e., the identity matrix, while $\bm{\Psi} = \int_0^1 \bm{\psi}(t) \bm{\psi}^\top(t) dt \neq \mathbb{1}$ for non-orthogonal basis functions such as $B$-spline basis. Throughout this study, we assume that $\bm{\Psi}$ is uniquely defined. In practice, the matrix of inner products between the basis functions, $\bm{\Psi}$, may be computed exactly. However, there may be some cases that $\bm{\Psi}$ is not unique. In such cases, the inner products can be obtained approximately, which can easily be done using the \texttt{inprod} function in the \texttt{R} package ``fda'' \citep{fdap}. Then, the optimization problem~\eqref{eq:cov} can be expressed as follows:
\begin{equation*}
\bm{w} = \arg \max \frac{\bm{w}^\top \bm{\Psi} \bm{\Sigma}_{\bm{D} Y} \bm{\Sigma}_{\bm{D} Y}^\top \bm{\Psi} \bm{w}}{\bm{w}^\top \bm{\Psi} \bm{w}}.
\end{equation*}
In what follows, the first eigenvector $\bm{w}^{(1)}$ is obtained by solving
\begin{equation}\label{eq:bsw1}
\bm{\Sigma}_{\bm{D} Y} \bm{\Sigma}_{\bm{D} Y}^\top \bm{\Psi} \bm{w}^{(1)} = \lambda_{\max} \bm{w}^{(1)}.
\end{equation}

Let us denote $\bm{\Psi} = \bm{\Psi}^{1/2} ( \bm{\Psi}^{1/2} )^\top$. Then, we have
\begin{align*}
 (\bm{w}^{(1)})^\top \bm{\Psi} \bm{w}^{(1)} &= (\bm{w}^{(h)})^\top \bm{\Psi}^{1/2} ( \bm{\Psi}^{1/2} )^\top \bm{w}^{(1)}, \\
&= (\widetilde{\bm{w}}^{(1)})^\top \widetilde{\bm{w}}^{(1)},
\end{align*} 
where $\widetilde{\bm{w}}^{(1)} = ( \bm{\Psi}^{1/2} )^\top \bm{w}^{(1)}$ and $\bm{w}^{(1)} = ( \Psi^{-1/2} )^\top \widetilde{\bm{w}}^{(1)}$. Accordingly, the problem~\eqref{eq:bsw1} can be expressed as follows:
\begin{equation*}
( \bm{\Psi}^{1/2} )^\top \bm{\Sigma}_{\bm{D} Y} \bm{\Sigma}_{\bm{D} Y}^\top \bm{\Psi}^{1/2} \widetilde{\bm{w}}^{(1)} = \lambda_{\max} \widetilde{\bm{w}}^{(1)}.
\end{equation*}
Then, the first PLS component is given by
\begin{equation*}
\bm{\xi}^{(1)} = \bm{A}  \widetilde{\bm{w}}^{(1)},
\end{equation*}
where $\bm{A} = \bm{D} ( \bm{\Psi}^{1/2} )^\top$. The subsequent PLS components are computed iteratively but with the basis expansion coefficients. Consequently, similar to univariate setting as in \cite{Aguilera2010, Aguilera2016}, the FPLS regression of $Y$ on $\bm{\X}(t)$ (i.e., Model~\eqref{eq:msof}) is equivalent to the PLS regression of $Y$ on $\bm{A}$ so that at each step of the PLS algorithm, both models produce the same PLS components \citep[see e.g.,][for more information]{Aguilera2019}. 

\section{The RFPLS method}\label{sec:app_B}

\subsection{M-estimate}\label{sec:3.1}

Before presenting the details of the proposed method, we summarize the idea of M estimation. Let us consider the standard regression problem between the scalar response $Y$ and basis expansion coefficients $\bm{A}$, i.e., $Y_i = \bm{A}_i \bm{\theta} + \epsilon_i$, where $\bm{\theta}$ denotes the vector of model parameters and $\epsilon_i$ is the error term. Then, the LS estimator of $\bm{\theta}$ is given by
\begin{equation*}
\widehat{\bm{\theta}} = \underset{\begin{subarray}{c}
  \bm{\theta}
  \end{subarray}}{\argmin} \sum_{i=1}^n ( Y_i - \bm{A}_i \bm{\theta} )^2.
\end{equation*}
However, in case of any departure from the model assumptions, such as normally distributed error terms or in the presence of outliers, the LS estimator produces biased estimates for the regression parameter. In such a case, the robust estimators, including the M-estimator, produce better estimates than the LS estimator. Let us consider one symmetric and nondecreasing loss function $\rho(\cdot)$ such as Tukey's bisquare loss function:
\begin{equation*} \label{eq:exploss}
        \rho(u)=
        \left\{ \begin{array}{ll}
            1 - [ 1 - ( \frac{u}{c} )^2 ]^3 & \text{if}~ \vert u \vert \leq c, \\
            1 & \text{if}~ \vert u \vert > c,
        \end{array} \right.
\end{equation*}
where $c$ is a tuning parameter. The M-estimator of $\bm{\theta}$ is then defined as follows:
\begin{equation*}
\widehat{\bm{\theta}}_{\rho} = \underset{\begin{subarray}{c}
  \bm{\theta}
  \end{subarray}}{\argmin} \sum_{i=1}^n \rho ( Y_i - \bm{A}_i \bm{\theta} ).
\end{equation*}
Note that if $\rho(u) = u^2$, then $\widehat{\bm{\theta}}_{\rho} = \widehat{\bm{\theta}}$.

Let $r_i^e = \rho(e_i) / e_i^2$, where $e_i = Y_i - \bm{A}_i \bm{\theta}$, denote the weight associated to $i^\textsuperscript{th}$ observation. Then, the M-estimator of $\bm{\theta}$ can also be expressed as a weighted LS estimator as follows:
\begin{equation}\label{eq:wei1}
\widehat{\bm{\theta}}_{\rho} = \underset{\begin{subarray}{c}
  \bm{\theta}
  \end{subarray}}{\argmin} \sum_{i=1}^n r_i^e ( Y_i - \bm{A}_i \bm{\theta} )^2.
\end{equation}
If the model assumptions hold and no outliers are present in the data, then all the observations take weight close to one, and the M- and LS estimators become similar. Otherwise, the observations having large residuals take weights close to zero, and M-estimator produces better estimates compared with LS by down-weighting the effects of outlying observations \citep[see, e.g.,][for more information]{Serneels2005}.

\subsection{The RFPLS}

Because of the infinite-dimensional nature of the functional predictors, as in FPLS, we consider approximating the RFPLS components and parameter function via a finite-dimensional basis function expansion of the functional predictors. Consider the PLS regression of $Y$ on the random vector $\bm{A} = \bm{D} ( \bm{\Psi}^{1/2} )^\top$, i.e.,
\begin{equation*}
Y = \bm{1} \gamma_0 + \widetilde{\bm{\xi}}^{(h)} \bm{\gamma}.
\end{equation*}
To obtain an estimate for $\bm{\beta}$ that is robust to both vertical outliers and leverage points, we consider multiplying the weight in~\eqref{eq:wei1} with a second weight $r^a_i$ for down-weighting the effects of outliers in the predictor space as follows:
\begin{equation*}
\widehat{\bm{\theta}}_{r} = \underset{\begin{subarray}{c}
  \bm{\theta}
  \end{subarray}}{\argmin} \sum_{i=1}^n r_i^e r^a_i ( Y_i - \bm{A}_i \bm{\theta} )^2.
\end{equation*}

For the PLS regression of $Y$ on $\bm{A}$, we first compute the robust eigenvectors $\widetilde{\bm{w}}_r$ via the weighted covariance, i.e., by optimizing the following objective function:
\begin{align*}
\widetilde{\bm{w}}_{r} &= \text{Cov}_{r}(Y, \bm{A}) \\
& \Leftrightarrow \underset{\begin{subarray}{c}
  \widetilde{\bm{w}}	
  \end{subarray}}{\argmin} \sum_{i=1}^n r_i (Y_i - \bm{A}_i \widetilde{\bm{w}} )^2,
\end{align*}
where $r_i = r_i^e r^a_i$. Then, the robust PLS components are computed as $\bm{\xi}_{r} = \bm{A} \widetilde{\bm{w}}_{r}$. After obtaining the estimate of $\bm{\gamma}$, $\bm{\widehat{\gamma}}_r$, the robust estimate of $\bm{\beta}$ is given by $\widehat{\bm{\beta}} = (\bm{\Psi}^{-1/2})^\top \widetilde{\bm{w}}_{r} \bm{\widehat{\gamma}}_r$. Note that the weighted PLS is a special case of the standard PLS regression. The quantities of the weighted PLS are the PLS quantities computed from the weighted observations, $\sqrt{r_i} \bm{A}_i$ and $\sqrt{r_i} Y_i$. Herein, the weights are not fixed and they need to be updated in each PLS iteration. Thus, the iteratively reweighted PLS (IRPLS) algorithm is used to obtain the final weighted PLS quantities.

In the IRPLS algorithm, the weights for $r_i^e$ is computed by $r_i^e = f (e_i / \widehat{\sigma})$, where $\widehat{\sigma}$ is the median absolute deviation estimate of residual scale, i.e., $\widehat{\sigma} = \underset{i}{\text{median}} \vert e_i - \underset{j}{\text{median}}~ e_j \vert$ and $f(\cdot)$ is the Hampel's loss function
\[ f(x) = \begin{cases}
x, & 0 \leq \vert x \vert \leq c_1 \\
c_1 \text{sign}(x), & c_1 \leq \vert x \vert \leq c_2 \\
\frac{c_1 (c_3 - \vert x \vert )}{c_3 - c_2} \text{sign}(x), & c_2 \leq \vert x \vert \leq c_3 \\
0, & c_3 \leq \vert x \vert 
\end{cases}
\]
where $c_1 = 1.65$, $c_2 = 1.96$ and $c_3 = 3.09$. The weights for $r_i^a$ that bounds the effects of outliers in the PLS components are computed as follows:
\begin{equation*}
r_i^a = f \left(\frac{\Vert \bm{\xi}_{ri} - \text{med}_{L_1} (\bm{\xi}_r) \Vert}{\text{median}_i \Vert \bm{\xi}_{ri} - \text{med}_{L_1} (\bm{\xi}_r) \Vert} \right),
\end{equation*}
where $\text{med}_{L_1} (\bm{\xi}_r) $ denotes the $L_1$ median computed from the collection of component vectors \citep[see e.g.,][]{Serneels2005}. In the first step of the IRPLS algorithm, the starting values for the weights $r_i^e$ are computed using $r_i^e = Y_i - \underset{j}{\text{median}}~Y_j$, while the starting values for the weights $r_i^a$ are computed by replacing component vectors with $\bm{A}_i$. In each step of the algorithm, the SIMPLS method of \cite{simpls} is applied to the weighted data matrices, i.e., $\sqrt{r} Y$ and $\sqrt{r} \bm{A}$ where $r = r^e r^a$, to obtain the PLS estimate $\widehat{\gamma}_r$ and PLS component matrix $\bm{\xi}_r$. In each step, the PLS components are corrected by dividing each row of $\bm{\xi}_r$ by $\sqrt{r_i}$ and the residuals are recomputed $r_i^e = Y_i - \bm{\xi}_{ri} \widehat{\gamma}_r$ to update the weights $r_i = r_i^e r_i^a$. For each step of the IRPLS algorithm, this process is repeated until convergence is achieved. Convergence is reached if the relative difference in norm between two PLS estimates $\widehat{\gamma}_r$ is smaller than a threshold value, e.g. $10^{-2}$.

Let us consider the approximate model of Model~\eqref{eq:msof} by the robustly obtained PLS components
\begin{equation*}
Y = \bm{1} \delta_{0} + \bm{\xi}_{r}^{(h)} \bm{\delta} = \bm{1} \delta_{0} + \bm{D} \bm{\Psi} \bm{\beta}.
\end{equation*}
For the final estimate of $\bm{\delta}$, we consider the M-estimator, i.e.,
\begin{equation*}
\widehat{\bm{\delta}}_{\rho} = \underset{\begin{subarray}{c}
  \bm{\delta}	
  \end{subarray}}{\argmin} \sum_{i=1}^n \rho (Y_i - \bm{\xi}_{ri}^{(h)} \bm{\delta} ),
\end{equation*}
where $\rho(\cdot)$ is the Tukey's bisquare loss function. For a given scale value $\mathcal{S}_n$, $\widehat{\bm{\delta}}_{\rho}$ is calculated by solving the following estimation functions:
\begin{equation} \label{eq:obj}
\vartheta(\bm{\delta}) := \sum_{i=1}^n \bm{\xi}_{ri}^{(h)} \kappa \left\lbrace \frac{Y_i - \bm{\xi}_{ri}^{(h)} \bm{\delta}}{\mathcal{S}_n} \right\rbrace = \bm{0},
\end{equation}
where $\kappa$ is the sub-gradient of $\rho(\cdot)$ up to constant $\frac{6}{c^2}$:
\begin{equation} \label{eq:psi}
\kappa(u)=
        \left\{ \begin{array}{ll}
            u [ 1 - ( \frac{u}{c} )^2 ]^2 & \text{if}~ \vert u \vert \leq c \\
            1 & \text{if}~ \vert u \vert > c
        \end{array} \right.
\end{equation}
Considering~\eqref{eq:obj} and~\eqref{eq:psi} together, $\vartheta(\bm{\delta})$ has the following form:
\begin{equation} \label{eq:ulast}
\vartheta(\bm{\delta}) = \sum_{i=1}^n \bm{\xi}_{ri}^{(h)} \omega_i e_i,
\end{equation}
where $\omega_i = \frac{\kappa(e_i)}{e_i}$ and $e_i = \frac{Y_i - \bm{\xi}_{ri}^{(h)} \widetilde{\bm{w}}_{\rho}}{\mathcal{S}_n}$. Finally, $\widehat{\bm{\delta}}_{\rho}$ is obtained by solving~\eqref{eq:ulast} as follows:
\begin{equation} \label{eq:rphi}
\widehat{\bm{\delta}}_{\rho} = \left( \sum_{i=1}^n \bm{\xi}_{ri}^{(h)} \omega_i (\bm{\xi}_{ri}^{(h)})^\top \right)^{-1} \left( \sum_{i=1}^n \bm{\xi}_{ri}^{(h)} \omega_i Y_i \right).
\end{equation}

In~\eqref{eq:rphi}, $\omega_i$ is a function of both $\bm{\delta}$ and $\mathcal{S}_n$. Thus, $\widehat{\bm{\delta}}_{\rho}$ is computed iteratively, where at iteration $l$, the previous $\widehat{\bm{\delta}}_{\rho}^{(l-1)}$ with the median absolute deviation estimator for $\mathcal{S}_n$, is used to estimate $\omega_i$ such that $\widehat{\omega}_i = \omega_i \vert_{\widehat{\bm{\delta}}_{\rho} = \widehat{\bm{\delta}}_{\rho}^{(l-1)}}$. In other words, $\widehat{\bm{\delta}}_{\rho}$ is computed via an iteratively reweighted least squares algorithm. The tuning parameter $c$ used in the bisquare loss function controls the degree of robustness for $\widehat{\bm{\delta}}_{\rho}$ and may have a significant effect on the efficiency and degrees of robustness \citep[see, e.g.,][]{Wang2007}. In practice, $c$ is chosen according to the desired robustness degree. On the other hand, the efficiency of $\widehat{\bm{\delta}}_{\rho}$ may vary with different choices of $c$. Therefore, we choose the tuning parameter $c$ based on the efficiency factor $\tau$ introduced by \cite{Wang2018} as follows:
\begin{equation*}
\widehat{\tau}(c) = \frac{\left[ \sum_{i=1}^n \frac{\partial \kappa(e_i + \delta)}{\partial \delta} \right]^2}{n \sum_{i=1}^n \kappa^2(e_i)},
\end{equation*}
where $\delta$ is a small constant. We consider the procedure based on the one proposed by \cite{Wang2007} to determine the tuning parameter $c$. The numerical analyses performed by \cite{Wang2007} showed that the number of iterations needed to compute $\widehat{\bm{\delta}}_{\rho}$ based on tuning parameter $c$ is small such as 1-3. The algorithm for the determination of the tuning parameter, which can easily be done by the  \Rlogo \ package \texttt{rlmDataDriven} \citep{rlmdd}, is as follows.

\begin{algorithm}[!htb]
Obtain the residuals $e_i(\widehat{\bm{\delta}}_{\rho 0})$ and median absolute deviation estimate for $S_n$, where $\widehat{\bm{\delta}}_{\rho 0}$ is the initial value computed based on LS estimator:
\begin{equation*}
\widehat{\bm{\delta}}_{\rho 0} = \arg~~\underset{\begin{subarray}{c}
  \bm{\delta}
  \end{subarray}}{\min} \sum_{i=1}^n [Y_i - (\bm{\xi}_{ri}^{(h)})^\top \bm{\delta} ]^2.
\end{equation*} 
Find the optimum $c$ value between 1 and 10 \citep{Wang2018}, leading to the largest efficiency factor $\widehat{\tau}(c)$.  \\
Compute $\widehat{\bm{\delta}}_{\rho}$ using M-estimator with the selected tuning parameter $c$ calculated in the previous step.
\caption{\small{Tuning parameter selection algorithm}}
\label{alg:tuning}
\end{algorithm}

Let $\widehat{\bm{\delta}}_{\rho}$ denote the final robust estimate of $\bm{\delta}$. Then, the robust basis expansion coefficients of the regression coefficient function $\bm{\beta}(t)$ is obtained by
\begin{equation*}
\widehat{\bm{\beta}}_{r}^{(h)} = ( \Psi^{-1/2} )^\top \bm{W}_{r}^{(h)} \widehat{\bm{\delta}}_{\rho}.
\end{equation*}
Finally, the robust estimate of the parameter functions $\widehat{\bm{\beta}}_{r}^{(h)}(t) = [ \widehat{\bm{\beta}}_{r 1}^{(h)}(t), \ldots, \widehat{\bm{\beta}}_{r M}^{(h)}(t) ]^\top$ are obtained as follows:
\begin{equation*}
\widehat{\bm{\beta}}_{r}^{(h)}(t) = \bm{\psi}(t) \widehat{\bm{\beta}}_{r}^{(h)}.
\end{equation*}

\end{document}